\documentclass[3p,times,12pt]{elsarticle}

\journal{Nuclear Physics B}

\usepackage[colorlinks]{hyperref}
\usepackage{amsmath,amssymb}
\allowdisplaybreaks[2]

\usepackage{epstopdf}

\newcommand{\beq}{\begin{equation}}
\newcommand{\eneq}{\end{equation}}
\newcommand{\upa}{\uparrow}
\newcommand{\dwa}{\downarrow}
\newcommand{\Upa}{\Uparrow}
\newcommand{\Dwa}{\Downarrow}

\newcommand{\nn}{\nonumber}
\newcommand{\h}{\mathcal{H}}
\newcommand{\hb}{\mathcal{H}_{\mathbf{B}}}

\newcommand{\nor}[1]{\::\!#1\!:\:}
\newcommand{\Sg}{\mathbf{S}_G}
\newcommand{\Z}{\mathcal{Z}}

\begin{document}

\begin{frontmatter}



\title{Enhanced Coherence of a Quantum Doublet Coupled to Tomonaga-Luttinger Liquid Leads }


\author[unipg]{Antonio Cirillo}
\ead{antonio.cirillo@fisica.unipg.it}
\author[unipg]{Matteo Mancini}
\ead{matteo.mancini@fisica.unipg.it}
\author[unical]{Domenico Giuliano\corref{cor1}}
\ead{domenico.giuliano@fis.unical.it}
\author[per,unipg]{Pasquale Sodano}
\ead{pasquale.sodano@pg.infn.it}

\address[unipg]{Dipartimento di Fisica, Universit\`{a} di
Perugia,   Via A. Pascoli, I-06123,
Perugia, Italy and \\
I.N.F.N., Sezione di Perugia, Via A. Pascoli, I-06123,
Perugia, Italy }
\address[unical]{Dipartimento di Fisica, Universit\`{a} della Calabria Arcavacata di Rende
I-87036, Cosenza, Italy and \\
I.N.F.N., Gruppo collegato di Cosenza, Arcavacata di Rende
I-87036, Cosenza, Italy}
\address[per]{Perimeter Institute for Theoretical Physics  31 Caroline St.
N,Waterloo ON, N2L 2Y5, Canada}
\cortext[cor1]{Corresponding author}

\begin{abstract}
We use boundary field theory to describe the phases accessible to
a tetrahedral qubit coupled to Josephson junction chains acting as
Tomonaga-Luttinger liquid leads. We prove that, in a pertinent
range of the fabrication and control parameters, an attractive
finite coupling fixed point emerges due to the geometry of the
composite Josephson junction network. We show that this new stable
phase is characterized by the emergence of a quantum doublet which
is robust not only against the noise in the external control
parameters (magnetic flux, gate voltage) but also against the
decoherence induced by the coupling of the tetrahedral qubit with
the superconducting leads. We provide protocols allowing to read
and to manipulate the state of the emerging quantum doublet and
argue that a tetrahedral Josephson junction network operating near
the new finite coupling fixed point may be fabricated with today'
s technologies.

\end{abstract}


\end{frontmatter}

\section{Introduction}
\label{sec:intro}

Quantum impurity models \cite{boundary} provide a natural paradigm
to describe a large number of nonperturbative phenomena occurring
in one dimensional quantum devices such as point contacts,
constrictions, crossed quantum wires and Josephson junction (JJ)
chains \cite{beenakker1,beenakker2,chamon,chamon2,giuso4}. While a
standard perturbative approach is accurate when the impurity is
weakly coupled to the environmental modes needed to fully describe
the quantum device, there are situations in which impurities are
strongly coupled to such environmental modes: when this happens, it
is impossible to disentangle the impurity from the rest of the
system, the perturbative approach breaks down and one has to
resort to the non perturbative tools provided by boundary field
theories (BFT) \cite{boundary,cardy}, which have been shown to
yield accurate descriptions of many realistic low dimensional
systems \cite{giamarchi}.

Prototypical non perturbative impurities states are realized in
systems exhibiting the Kondo effect
\cite{tsvel,schlott,nozi,afflud}, or in situations where static
defects appear in Tomonaga-Luttinger liquids (TLL)s; for both
settings, a renormalization group approach leads, after
bosonization \cite{giamarchi}, to the emergence of boundary
sine-Gordon models\cite{fendlud,chang}. In both cases, the
interaction with the impurity makes the boundary coupling strength
to scale to a stable strongly coupled fixed point (SFP) which is
characterized by a fully screened spin in Kondo systems or by the
effective disappearance of the impurity in a ``healed'' TLL
\cite{kanefish1,kanefish2}. More remarkable states are achieved
when a finite coupling fixed point (FFP) - characterized by new
non-trivial universal indices - emerges; for instance, this happens in the
overscreened Kondo problems \cite{tsvel,schlott,nozi,afflud}, or
in crossed TLLs where, as a result of the crossing, some operators
turn from irrelevant to marginal, leading to correlation functions
exhibiting power-law decays with nonuniversal exponents
\cite{reyes,chamon,chamon2}.

Quantum impurities are realizable also in superconducting
Josephson devices \cite{giuso1,giuso2,giusoepl,giusolast}.
Superconducting Josephson chains with a weak link
\cite{glark,giuso1} and SQUIDs \cite{glhek,giuso2} may be indeed
described by boundary sine Gordon models yielding a phase diagram
with only two fixed points: an unstable weakly coupled fixed point
(WFP), and a stable one at strong coupling. At variance, for
pertinent architectures of the Josephson junction network (JJN)
one may find a range of fabrication and control parameters where a
stable FFP emerges in the phase diagram \cite{giuso4} allowing for
the engineering of superconducting devices exhibiting enhanced
quantum coherence \cite{giusonew} and $4e$ superconductivity
\cite{giusoepl}. The stable fixed point is associated with the
emergence of a doubly degenerate ground state which may be
regarded as a quantum doublet described by a spin $1/2$ degree of
freedom coupled to the plasmon modes via the boundary interaction;
as a result, one may use superconducting devices not only as good
candidates for the design of solid state quantum bits \cite{shon0}
but also as efficient quantum simulators of the various physical
behaviors realizable in Kondo systems.

\begin{figure}[htb]
\begin{center}
    \includegraphics[width=0.3\textwidth]{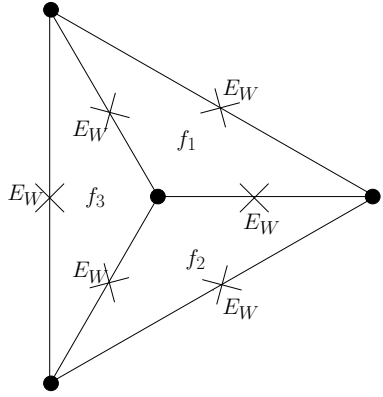}
\end{center}
 \caption{The central region ${\bf T}$.
The junctions are assumed to have the same Josephson energy $E_W$
while the three loops are threaded by the dimensionless fluxes
$f_1 , f_2 , f_3$, respectively.}
  \label{fig:device2}
\end{figure}

Engineering quantum doublets robust against noise and decoherence
is of paramount importance for applications to quantum information
processing. Quite recently a remarkably robust two-level quantum
system has been shown to emerge in a device made with six JJs
arranged (see Fig.\ref{fig:device2}) in a symmetric tetrahedral
geometry \cite{blatter1,blatter2}. In
Refs.\cite{blatter1,blatter2} it has been pointed out that, when
each internal loop is pierced by a dimensionless magnetic flux
$f=\pi$, the ground state is doubly degenerate and that the
degeneracy is robust against small variations in the applied gate
voltages and/or in the applied magnetic flux, since the degeneracy
is split only to second order in the charge and flux noise.
Remarkably, the design of a tetrahedral quantum bit
\cite{blatter1,blatter2} may be modified so as to make it robust
also against the noise in the Josephson junction energy
\cite{Usma}.

For any practical purpose (control or reading out the state of the
quantum bit) one needs to connect the tetrahedral quantum bit to
external leads, which may induce new decoherence effects spoiling
the coherence of the quantum doublet \cite{wiring}.  As we shall
show, realizing the leads with TLLs, enhances the coherence of
this device. This happens if one uses one-dimensional
superconducting leads which may be mapped onto TLLs with the
Luttinger parameter $g$ depending on the fabrication parameters of
the JJN \cite{glhek,glark,giuso1,giuso2,shulz,yang}. In this
paper, we address this issue by analysing the device made by the
central tetrahedral JJN - ${\bf T}$ - depicted in
Fig.\ref{fig:device2} connected to three JJ chains ending in bulk
superconductors at fixed phase $\phi_j, j=1,2,3$; the resulting
network is depicted in Fig.\ref{fig:device}. We show explicitly
that the interaction of the central region ${\bf T}$ with the
low-energy collective excitations (plasmons) of the leads merely
renormalizes the ``bare'' parameters of ${\bf T}$ and, thus, does
not break the tetrahedral ``symmetry'' responsible for the robust
groundstate degeneracy\cite{blatter1,blatter2}. As a result we
exhibit a JJN which is robust also against fluctuations of the
lead's parameters and argue that such network may be fabricated
with nowadays technolgies \cite{haviland}.

The paper is organized as follows.

In section \ref{device_0}, we derive an effective spin-$1/2$
Hamiltonian for  ${\bf T}$ (Fig.\ref{fig:device2}). In agreement
with the results of refs.\cite{blatter1,blatter2}, we find that
the spectrum of the effective Hamiltonian admits a twofold
degenerate groundstate and argue that, in the effective theory,
${\bf T}$ may be regarded as a spin-$1/2$ degree of freedom ${\bf
S}_G$. Finally, we derive the low-energy description of the JJN
depicted in Fig.\ref{fig:device}. We show then that the effective
theory is a 1+1 dimensional field theory, where the central region
is described by a spin interacting with the leads via a pertinent
boundary interaction.

In section \ref{weak0} we use a perturbative approach to account
for the couplings between ${\bf T}$ and the TLL leads realized
with JJ chains. Here, we use a renormalization group approach to
determine the flow of the running boundary coupling strengths, as
a function of the system size and we determine the range of values
of $g$ and $f$ where the perturbative approach breaks down.

In section \ref{strong0}, we analyze the large-scale behavior of
the JJN depicted in Fig.\ref{fig:device} in the strong coupling
limit where the boundary effects induced by
${\bf T}$ become relevant. In particular, we find the set of
minima of the boundary interaction potential at the SFP and
construct the instanton operators connecting two degenerate
minima. Using the renormalization group approach
\cite{Cardy1996scaling}, we analyse the flow of the coupling
constants associated to the instantons. Finally, we show that -
for $-\frac{\pi}{10} \lesssim f-\pi \lesssim \frac{\pi}{10}$ and
$1 < g < 3$ - a finite coupling fixed point emerges.

In section \ref{emerge0} we show that instantons are responsible
for the emergence of a two-level quantum system which, due to the
network's architecture, is robust, for $-\frac{\pi}{10} \lesssim
f-\pi \lesssim \frac{\pi}{10}$, against the decoherence arising
from coupling the cental region to the leads. Moreover, we show
how the state of the emerging quantum doublet may be manipulated
by acting upon the external control parameters.

In section \ref{engineer0} we point out that the quantum doublet
emerging in a tetrahedral JJN operating near the FFP may be
realized with today's technologies.

Section \ref{conclusions} is devoted to concluding remarks.

 The appendices provide the necessary mathematical
background to follow the analysis carried in the text.

\section{Boundary field theory description of the tetrahedral JJN}
\label{device_0}

\begin{figure}[htb]
\begin{center}
    \includegraphics[width=0.9\textwidth]{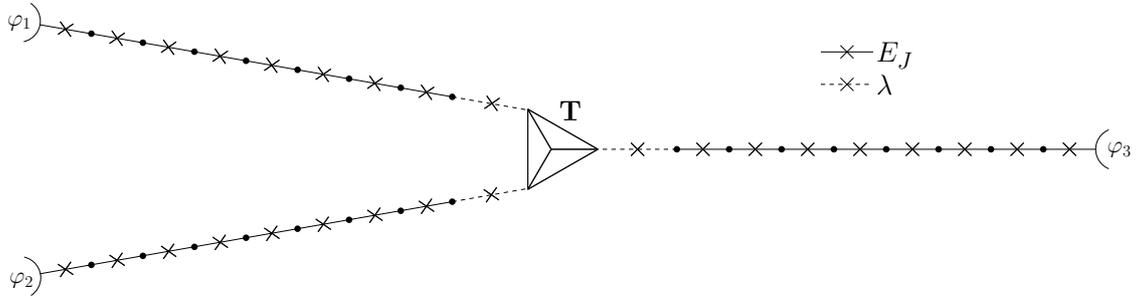}
\end{center}
  \caption{The tetrahedral JJN. The central region {\bf T}, depicted in Fig.\ref{fig:device2}, is
made with four superconducting grains connected by six quantum
Josephson junctions tuned nearby the degeneracy between two charge
eigenstates. {\bf T} is connected, via a coupling $\lambda\ll
E_W$, to three leads realized with one-dimensional Josephson
junction arrays fabricated with junctions of nominal Josephson
energy $E_J$ and ending in three bulk superconductors held at
fixed phases $\varphi_1 , \varphi_2 , \varphi_3$}.
  \label{fig:device}
\end{figure}

In this section, we derive the low-energy, long-wavelength
description of the JJN depicted in Fig.\ref{fig:device}. We shall
find that this JJN may be described by a 1+1 dimensional field
theory, with a pertinent boundary interaction describing the
central region. In the following we shall firstly derive the
Hamiltonian associated to the central region {\bf T} and then
construct the BFT describing the JJN in which {\bf T} is connected
to TLL leads realized with JJ chains ending with three bulk
superconductors held at fixed phases $\varphi_1 , \varphi_2 ,
\varphi_3$.

\subsection{The central region {\bf T}}
\label{device_1}

${\bf T}$ is fabricated with six quantum Josephson junction,
joined to each other as depicted in Fig.\ref{fig:device2}. We
assume that, for each junction, the charging energy $E_C$ is much
bigger than the Josephson energy $E_W$. To prevent Coulomb
blockade from forbidding charge transport across {\bf T}, we
further assume that to each superconducting grain is applied a
gate voltage $V_g$, tuned nearby the degeneracy between the charge
eigenstates with charge equal to ${\mathcal N}$ and ${\mathcal
N}+1$. Under these assumptions  one can describe each
superconducting grain with a quantum spin-$1/2$ variable ${\bf
S}_0^{(i)}$ \cite{giuso1,giuso2,giusolast}; this leads to an
effective spin-$1/2$ representation for the Hamiltonian
$\mathcal{H}_{\bf T}$ describing the central region.

If each internal loop in Fig.\ref{fig:device2} is pierced by a
(dimensionless) magnetic flux $f$ (i.e. $f_1 = f_2 = f_3 = f$), a
standard procedure \cite{giusolast} yields
\begin{equation}
  \begin{split}
    \mathcal{H}_{\bf T} &= -H\sum_{i=0}^{3}(S_0^{(i)})^z
    -\frac{E_W}{2} \left[ \sum_{i=1}^3\left( e^{i f}(S_0^{(i)})^{+}( S_0^{(i+1)})^{-}+
    e^{-i f}(S_0^{(i+1)})^{+}(
   S_0^{(i)})^{-} \right)\right.+ \\
    &+ \left. \sum_{i=1}^3\left( (S_0^{(i)})^{+}( S_0^{(0)})^{-}+(S_0^{(0)})^{+}( S_0^{(i)})^{-}\right) \right]
   \:\:\:\: .
 \end{split}
 \label{tetra1}
\end{equation}
\noindent In Eq.(\ref{tetra1}), ${\bf S}_0^{(i)}$, $i=1,2,3$,
denotes an effective spin-$1/2$ operator lying on the three outer
sites of ${\bf T}$, while
 ${\bf S}_0^{(0)}$ lies on the central site. The parameter $H \propto e^* V_g - N - \frac{1}{2}$ is determined by the uniform
gate voltage bias.

$\mathcal{H}_{\bf T} $ can be exactly diagonalized: its
eigenvalues and eigenstates are given in \ref{appe1}. There we
show also that, for $\frac{\pi}{2} < f < \frac{3 \pi}{2}$ and $| H
| \ll E_W$, the ground state is twofold degenerate; here, we denote
the states of the degenerate quantum doublet as $ \lvert 0 , 1 \rangle
\equiv \lvert \Uparrow \rangle $ and $ \lvert 0 , 2 \rangle =
\lvert \Downarrow \rangle$. The effective spin-$1/2$ operator
${\bf S}_G$, acting onto the two-dimensional subspace spanned by
$\{\lvert \Uparrow \rangle ,\lvert \Downarrow \rangle \} $ may be
then represented as
\begin{equation}
{\bf S}_G^a = \frac{1}{2} \: \sum_{ \sigma , \sigma^{'}} \: |
\sigma \rangle \langle \sigma^{'} | \: \tau_{\sigma ,
\sigma^{'}}^a \:\:\:\: , \label{state6}
\end{equation}
where $a=1 \dots 3$, $\tau$ are the Pauli matrices and $\sigma =
\Uparrow , \Downarrow$.

Although a spin-$1/2$ degree of freedom emerges quite naturally in
pertinently engineered JJN \cite{giuso4,giusoepl,giusolast}, its
robustness - against any detuning of $e^* V_g$ off the degeneracy
value $N + \frac{1}{2}$ as well as against any small deviation of
the dimensionless magnetic flux $f$ from its optimal value $\pi$ -
is a very challenging task. Indeed in all the devices analyzed in
Refs.\cite{giuso4,giusoepl,giusolast} the twofold degeneracy is
realized only if $f$ is fine-tuned to $\pi$, since any
displacement from $f=\pi$ breaks the doublet degeneracy already to
the first order in $f-\pi$. At variance, due to the frustration
induced by the presence of the central spin $S_0^ (0)$, the
tetrahedral central region {\bf T} is much more stable against
noise in the external control parameters.

\subsection{The emerging quantum doublet ${\bf S}_G$}
\label{device_2}

A simple symmetry argument provides us with an hint on why the
architecture of the central region is relevant for the robustness
of the emerging quantum doublet. For this porpouse we first prove
that this robustness depends crucially on if the number of
junctions needed to fabricate the central region is even or odd.

If the central region is realized with an odd number of junctions
-as it happens, for instance, in the Y-shaped networks analyzed in
\cite{giuso4}- the ground state degeneracy is a consequence of the
invariance of the effective Hamiltonian under time-reversal
symmetry $\hat{T}$ which, for a three-spin effective Hamiltonian
describing a triangle-shaped central region \cite{giuso4}, is
explicitly realized as

\begin{equation}
  \hat{T} = [ \prod_{ j = 1}^3 \sigma_j^x ] \hat{T}_{123}
\;\;\;\; , \label{symme1}
\end{equation}
\noindent where the operator $[ \prod_{ j = 1}^3 \sigma_j^x ] $
($\sigma^x $ being the first Pauli matrix) changes the sign of
each one of the three spins, while $\hat{T}_{123}$ reverses the
label order of the spins (that is, $(123) \longrightarrow (321)$
\footnote{Notice that multiplication of $[ \prod_{ j = 1}^3
\sigma_j^x ]$ by  $\hat{T}_{123}$ is needed in order for the
symmetry to be preserved also when a magnetic flux is applied}).
Since, when the number of spins is odd, the ground state has total
spin 1/2, and since $\hat{T}$ reverses the sign of the total spin,
one sees immediately that the ground state must be twofold
degenerate. Such a degeneracy is, however, easily spoiled by a
displacement of the applied gate voltage off the degeneracy point,
i.e., by a nonzero value of the parameter $H$, which is coupled
linearly to the total spin of the state: any detuning of $e^* V_g$
off the degenerate value $N + \frac{1}{2}$ breaks then the ground
state degeneracy  already to the first order in $e^* V_g - N -
\frac{1}{2}$. At variance, when the central region is effectively
described by a spin Hamiltonian with an even number of spins
(i.e., the number of junctions needed to fabricate the central
region is even) the ground state is a spin singlet and, thus,
insensitive to a nonzero value of $H$. Although robustness versus
accidental displacements in the applied gate voltage is
guaranteed, to get a degenerate ground state requires a fine
tuning of the applied flux $f$ to $f = \pi$: any displacement of
the applied flux off the optimal value $f - \pi$ breaks the
degeneracy between the two ground states already to the first
order in $f - \pi$.

The tetrahedral geometry of the central region $\h_{\bf T}$
provides an optimal compromise between the above complementary
issues, since the twofold ground state degeneracy is protected by
symmetries that are realized for any value of $f$. Indeed,
time-reversal $\hat{T}$ is now realized as ${\bf S}_s = {\bf I}_S
\hat{T}_{123} $ where ${\bf I}_S$ reverses the sign of all the
spins and $\hat{T}_{123} $ is defined in Eq.(\ref{symme1}).
Looking at the explicit form of the states $ | \Uparrow \rangle $
, $ | \Downarrow \rangle$, given in appendix \ref{appe1},  one may
readily see that

\beq | \Downarrow \rangle = {\bf I}_S | \Uparrow \rangle \;\;\; ,
\;\; | \Uparrow \rangle = {\bf I}_S | \Downarrow \rangle \;\;\;\;
, \label{symme2} \eneq \noindent and that \beq \hat{T}_{123} |
\Uparrow \rangle = e^{ - \frac{2 i \pi}{3}} | \Uparrow \rangle
\;\;\; , \;\; \hat{T}_{123} | \Downarrow \rangle = e^{ \frac{2 i
\pi}{3}} | \Downarrow \rangle \:\:\:\: . \label{symme3} \eneq
\noindent Since  ${\bf I}_S$ is not a symmetry of $\h_{\bf T}$,
one concludes that   ${\bf I}_S | \Uparrow  \rangle $ must be an
eigenstate of $\h_{\bf T}$, independent of  $| \Uparrow  \rangle$,
but  degenerate in energy with it, as explicitly shown in
\ref{appe1}.  This ensures that the twofold ground state
degeneracy is allowed within a rather large range of values for
the applied flux $f$.

Robustness against variations in the external flux may be easily
seen also with the help of a perturbative analysis.Indeed, in
order to break the degeneracy between $ | \Uparrow \rangle , |
\Downarrow \rangle$ one may look at situation where the three
fluxes are split according to $f_j \to f + \delta_j$ (with $|
\delta_j / f | \ll 1$). Writing $\h_{\bf T} [ \{ f + \delta_i \} ]
$ as
\begin{equation}
\h_{\bf T} [ \{ f + \delta_i \} ] \approx \h_{\bf T} [ \{ f  \} ]
- i \frac{E_W}{2} e^{ i f} \: \sum_{ i = 1}^3 [ \delta_i
(S_0^{(i)})^{+}( S_0^{(i+1)})^{-} ] + { \rm h.c.} + {\mathcal O} [
\{ \delta_i^2 \} ] \;\;\;\; , \label{secondo1}
\end{equation}\noindent
one may easily compute matrix elements of $\h_{\bf T} [ \{ f +
\delta_i \} ] - \h_{\bf T} [ \{ f \} ]$ within the subspace
spanned by $ | \Uparrow \rangle , | \Downarrow \rangle$, provided
one knows that:
\begin{subequations}
\begin{align}
  \langle \sigma | \sum_{ i = 1}^3
  [ \delta_i (S_0^{(i)})^{+}( S_0^{(i+1)})^{-}] | \sigma \rangle
  &\propto \sin ( f )  \,\sum_{ i = 1}^3 \delta_i \\
  \langle \sigma | \sum_{ i = 1}^3
  [ \delta_i (S_0^{(i)})^{+}( S_0^{(i+1)})^{-} ] | \bar{\sigma} \rangle
  &\propto \sin ( f ) \,\left( e^{i\frac{\pi}{3}}\delta_1 - \delta_2 + e^{-i\frac{\pi}{3}}\delta_3 \right)
\;\;\;\; ,
\end{align}
\label{secondo2}
\end{subequations}\noindent
with $\sigma = \Uparrow , \Downarrow$, and $\bar{\Uparrow} =
\Downarrow$, $ \bar{\Downarrow} = \Uparrow$. From
Eqs.(\ref{secondo2}), one sees that, for $f=\pi$, the ground state
stays degenerate at least up to first order in the displacements
$\{ \delta_j \}$ $\delta_j$. To   second order in the   $\{
\delta_j \}$, one finds that the low energy eigenstates $|
\Uparrow ' \rangle$ and $ | \Downarrow ' \rangle$ and their
corresponding eigenvalues are given by:
\begin{align}
    \left\lvert \Uparrow'\right\rangle &= \frac{1}{\sqrt{2}} \left( \left\lvert \Uparrow\right\rangle +
    e^{i\xi}\left\lvert \Downarrow\right\rangle \right) \nonumber \\
    \left\lvert \Downarrow'\right\rangle &= \frac{1}{\sqrt{2}} \left( \left\lvert \Uparrow\right\rangle -
    e^{i\xi}\left\lvert \Downarrow\right\rangle \right) \nonumber \\
    \varepsilon_{\Uparrow'} &= -E_{W}\left(1+\frac{\delta_1^2+\delta_2^2+\delta_3^2}{12}\right)+\Delta \nonumber \\
    \varepsilon_{\Downarrow'} &= -E_{W}\left(1+\frac{\delta_1^2+\delta_2^2+\delta_3^2}{12}\right)-\Delta
  \label{eq:en_split}
\;\;\;\; ,
\end{align}
where
\begin{align}
  e^{i\xi} &= \frac{e^{i\frac{2\pi}{3}}\delta_1^2 +\delta_2^2 + e^{-i\frac{2\pi}{3}}\delta_3^2}
  {\sqrt{(\delta_1^2-\delta_2^2)+(\delta_2^2-\delta_3^2)+(\delta_3^2-\delta_1^2)}} \nonumber \\
  \Delta&=\frac{E_W}{12} \sqrt{2}\sqrt{(\delta_1^2-\delta_2^2)^2 +(\delta_2^2-\delta_3^2)^2+(\delta_3^2-\delta_1^2)^2}
  \label{eq:const_split}
\:\:\:\: .
\end{align}
From Eq.(\ref{eq:en_split},\ref{eq:const_split}) one sees that the
second order corrections to $  \h[\{f\}]$ may be recasted into an
effective Hamiltonian $\mathcal{H}_\perp$,   describing the
central region as a transverse magnetic field term, given by
\begin{equation}
\mathcal{H}_\perp  = B_\perp \cos( \xi ) {\bf S}_G^x + B_\perp \sin( \xi ) {\bf S}_G^y \;\;\; , \;\; (B_\perp = -\Delta)
\;\;\;\; .
\label{perp}
\end{equation}\noindent
Of course, when all the $\delta_j$ are equal then $B_\perp=0$.

In the next section we shall derive the BFT describing the central
region ${\bf T}$ connected to three TLL leads realized with three
JJ chains fabricated with junctions of nominal Josephson energy
$E_J$ and ending in three bulk superconductors held at fixed
phases $\varphi_1 , \varphi_2 , \varphi_3$.

\subsection{Connecting {\bf T} to TLL leads: the boundary Hamiltonian.} \label{device_3}

To derive the BFT describing the tetrahedral JJN depicted in
Fig.\ref{fig:device} we require at first that the leads are
realized by one-dimensional JJN for which the Josephson energy
$E_J$ is bigger than the charging energy $E_C$; we further assume
that there is an uniform gate voltage $V_g$ acting on each
junction tuned nearby the degeneracy point between the states with
${\mathcal N}$, or ${\mathcal N} + 1$ Cooper pairs at each
junction. As a result \cite{giuso1}, each lead may be described by
a one-dimentional spin-$1/2$ chain; if each chain is made out of
$L$ sites, the Hamiltonian describing the leads is given by

\begin{equation}
\h_{\rm Leads} = - \frac{E_J}{2} \sum_{ a =1,2,3} \sum_{ j = 0 ,
L-1} \{ \sigma_{ a , j}^+ \sigma_{ a , j+1}^- + \sigma_{ a ,
j+1}^+ \sigma_{ a , j}^- \} + E^z \sum_{ a = 1,2,3} \sum_{ j =
0}^{L-1} \sigma_{ a , j}^z \sigma_{ a , j+1}^z \:\:\:\:
\label{lead1}
\end{equation}
\noindent with
\begin{align}
{\bf \sigma}_{a , j}^+ &= {\bf P}_G e^{ i \phi_{a , j}} {\bf
P}_G^\dagger \\
{\bf \sigma}_{a , j}^z &= {\bf P}_G \left[ - i\frac{
\partial }{
\partial \phi_{a , j}} -  V_g \right]{\bf P}_G^\dagger.
\end{align}
Here, $\phi_{a,j}$ is the is the phase of the superconducting
order parameter at site-$j$ of the $a$-chain, ${\bf P}_G$ is the
projector onto the subspace of the Hilbert space with either
${\mathcal N}$, or ${\mathcal N} + 1$ Cooper pairs at each
superconducting grain, and $E^z$ is the effective strength of the
charge interaction between nearest-neighboring junctions. As a
result the low-energy long wavelength limit of the Hamiltonian
\eqref{lead1} can be described by a one-dimensional spinless
TLL Hamiltonian, given by

\begin{equation}
\h_{\rm LL} = \frac{g}{4 \pi} \: \sum_{ a = 1}^3 \: \int_0^L \; d x \;
\left[ \frac{1}{u} \left( \frac{\partial \Phi_a}{ \partial t} \right)^2 +
u \left( \frac{\partial \Phi_a}{ \partial x} \right)^2  \right]
\;\;\;\; ,
\label{llut}
\end{equation}
\noindent where $\Phi_a$ describes the collective plasmon modes of
the leads, while the Luttinger parameters $g$ and $u$ are given by
$g = \frac{\pi}{2 ( \pi - {\rm arccos} ( \frac{\Delta}{2} ))}$, $u
= a E_J \left[ \frac{\pi}{2} \frac{\sqrt{1 -
(\frac{\Delta}{2})^2}}{ {\rm arccos ( \frac{\Delta}{2} )} }
\right]$ ($\Delta = (E^z - 3 E_J^2 / 16 E_c) / E_J$, $a$ is the
lattice step) \cite{glark,giuso1}.

The leads are connected to  {\bf T} by means of
three Josephson junctions, of nominal strength $\lambda \ll E_W <
E_J$, connecting the endpoints of the leads to the outer sites of
{\bf T}. The Hamiltonian describing this interaction is given by
\cite{giuso4}

\begin{equation}
  \mathcal{H}_\lambda= - \lambda\sum_{i=1}^3\left( (S_0^{(i)})^{+} e^{-i\frac{\Phi_{i}(0)}{\sqrt{2}}}+
  (S_0^{(i)})^{-} e^{i\frac{\Phi_{i}(0)}{\sqrt{2}}} \right)
  \label{eq:ht}
\;\;\;\; .
\end{equation} \noindent
Using Eq.(\ref{eq:ht}) together with the spectrum of {\bf T} given
in \ref{appe1}, the Schrieffer-Wolff (SW) procedure yields an
effective boundary Hamiltonian involving only the low-energy
degrees of freedom of {\bf T}. A rather lengthy computation yields

\begin{align}
  \hb =& 2 E_1 I \sum_{i}\cos\left[ \frac{\Phi_i(0)-\Phi_{i+1}(0)}{\sqrt{2}} \right] +
  4 E_z \mathbf{S}_G^z \sum_{i}\cos\left[ \frac{\Phi_i(0)-\Phi_{i+1}(0)}{\sqrt{2}} + \frac{\pi}{2}\right] +\nn\\
& + 4 E_3 \mathbf{S}_G^{x}\sum_{j} \cos\left[ \frac{2}{3}\pi(j-2) \right]\cos\left[ \frac{\Phi_j(0)-\Phi_{j+1}(0)}{\sqrt{2}}\right] + \nn \\
& + 4 E_3 \mathbf{S}_G^{y}\sum_{j} \sin\left[ \frac{2}{3}\pi(j-2)
\right]\cos\left[ \frac{\Phi_j(0)-\Phi_{j+1}(0)}{\sqrt{2}}\right]
+ B_{\parallel} \mathbf{S}_G^z \;\;\;\; ,
  \label{eq:hb2}
\end{align}
\noindent where
\begin{equation}
  E_1 = \frac{E_J \lambda ^2}{3 \left(E_J^2-4 H^2\right)} \:\: , \:
  E_z = \frac{2 H \lambda ^2}{\sqrt{3} \left(E_J^2 - 4 H^2\right)} \;\; , \;
  E_3 = 2 E_1
\;\;\; ,
  \label{eq:couplings}
\end{equation}
\noindent
and
\begin{equation}
B_\parallel = -\frac{24 \sqrt{3} \lambda^2 H (f-\pi)}{E_J^2}
\:\:\:\: .
\label{parallel}
\end{equation}
\noindent One sees from (\ref{eq:hb2}, \ref{eq:couplings},
\ref{parallel}) that the term $B_\parallel \sigma_G^z$ explicitly
breaks - to the second order in the control parameters $g$ and
$f$- the degeneracy between $ | \Uparrow \rangle$ and $ |
\Downarrow \rangle$. In the following sections we shall show how
the low-energy plasmon modes of the leads renormalize the
parameters of  $\hb$.

\section{Perturbative analysis near the WFP }
\label{weak0}

In this section, we determine the flow of the running boundary
coupling strengths, and argue about the emergence of
nonperturbative fixed point in the phase diagram accessible to the
tetrahedral JJN.

\subsection{Renormalization group flow of the boundary coupling near the WFP.} \label{weak2}

To check the stability of WFP, we derive the renormalization group (RG)  equations for the
running boundary coupling strengths. To do so,  we use of a  boundary
version of the RG approach to perturbed conformal field theories,
developed by Cardy \cite{cardy0}. The starting point is given by
the Euclidean boundary action  $S_\textbf{B}^{(I)}$, corresponding to the
Hamiltonain in  Eq.\eqref{eq:hb2}, which is given by

\begin{align}
S_\textbf{B}^{(I)} =& 2 E_1 \int_0^\beta d\tau \:  \sum_{j}\cos\left( \vec{\alpha}_j \cdot \vec{\chi}
(\tau)\right) + 4 E_z \int_0^\beta d\tau \: {\bf S}_G^z \sum_{j} \cos\left( \vec{\alpha}_j
\cdot \vec{\chi} (\tau)+\frac{\pi}{2}\right) +\nn \\
& + 4 E_3 \int_0^\beta d\tau \: {\bf S}_G^x \sum_{j}
\cos\left(\frac{2}{3}\pi(j-2) \right)\cos\left( \vec{\alpha}_j \cdot \vec{\chi} (\tau)\right)  +
4 E_3  \int_0^\beta  d\tau \:
{\bf S}_G^y \sum_{j} \sin\left(\frac{2}{3}\pi(j-2) \right)
\cos\left( \vec{\alpha}_j \cdot \vec{\chi} (\tau)\right)
\:\:\:\: ,
  \label{eq:hbfermion}
\end{align}
\noindent
with $\beta=(k_B T)^{-1}$,
\begin{equation}
\vec{\alpha}_1 = \begin{pmatrix}
1 \\ 0 \end{pmatrix} \;\;\; , \;\;
\vec{\alpha}_2 = \begin{pmatrix}
 - 1/2 \\ \sqrt{3} / 2 \end{pmatrix}
\;\;\; , \;\;
\vec{\alpha}_3 = \begin{pmatrix}
 - 1 /2  \\ - \sqrt{3} / 2  \end{pmatrix}
\:\:\:\: .
\label{definition}
\end{equation}\noindent
We have defined  $\vec{\chi} ( \tau ) = \vec{\chi} ( 0 , \tau ) =
[ \chi_1 ( x , \tau  ) , \chi_2 ( x , \tau  )]$, with  $\chi_1 ( x
, \tau  )= \frac{1}{ \sqrt{2}} [ \Phi_1 ( x , \tau ) -  \Phi_2 ( x
, \tau )  ]$, $\chi_2 ( x , \tau  )= \frac{1}{ \sqrt{6}} [ \Phi_1
( x , \tau ) +
 \Phi_2 ( x , \tau )  - 2\Phi_3 ( x , \tau )  ]$,
in order to evidence that the ``center of mass'' field
$\Phi ( x , \tau ) = \frac{1}{ \sqrt{3}} [ \Phi_1 ( x , \tau ) +
\Phi_2 ( x , \tau ) + \Phi_3 ( x , \tau ) ] $ decouples from
$S_\textbf{B}^{(I)} $ as expected from charge conservation at {\bf
T} \cite{chamon,chamon2,giuso4}. One may then write the ``free''
action $ S_0 = S_{\rm Lead}[ \vec{\chi} ]  + S_{\rm S}[ \Phi ,
\Theta ] $ only in term of the fields $ \chi_1 $ and $ \chi_2$.
Namely,

\begin{equation}
S_{\rm Lead} [ \vec{\chi} ]
=  \frac{g}{ 4 \pi}
\: \int_0^\beta \: d \tau \: \int_0^L \: d x \: \left[ \frac{1}{u} \left( \frac{
\partial \vec{\chi}  }{ \partial \tau} \right)^2 + u  \left( \frac{
\partial \vec{\chi} }{ \partial x} \right)^2 \right]
\:\:\:\: ,
\label{aclutt}
\end{equation}
\noindent and \beq S_{\rm S} [ \Phi , \Theta ] = - \frac{i}{2} \:
\int_0^\beta \: d \tau \: \frac{ d \Phi ( \tau )}{ d \tau } [ 1 -
\cos ( \Theta ( \tau )) ]  - \frac{1}{2} \: \int_0^\beta \: d \tau
\: \{ B_\parallel \cos ( \Theta ( \tau )) + B_\perp \sin ( \Theta
( \tau )) \cos ( \Phi ( \tau ) - \xi ) \} \:\:\:\: ,
\label{acspin} \eneq \noindent being the imaginary time action for
the quantum spin variable ${\bf S}_G$ reported in appendix
\ref{appe2} ($\Theta , \Phi$ are the polar angles: see appendix
\ref{appe2} for details). The partition function for the JJN,
${\mathcal Z}$, is then given by

\begin{equation}
{\mathcal Z}={\mathcal Z}_0 \langle {\bf T}_\tau e^{-S_{\bf
B}^{(I)}} \rangle_{(0)} \;\;\;\; . \label{partf1}
\end{equation}
\noindent where
\begin{equation}
{\mathcal Z}_0  = \int \: \prod_{ i = 1,2} {\mathcal D} \chi \:
\int {\mathcal D} \Omega e^{ -  S_{\rm Lead} [ \vec{\chi} ] -
S_{\rm S} [ \Phi , \Theta ] } \;\;\;\; . \label{parf2}
\end{equation}
\noindent ${\bf T}_\tau$ denotes the imaginary time time-ordered
product, and the boundary interaction action is written as

\beq
 S_\textbf{B}^{(I)} \equiv  \sum_{\vec\alpha \in \mathbb{A}}\sum_{a=x,y,z}
g_{\vec\alpha}^a \int_0^\beta \frac{d\tau}{L} \:
\nor{e^{i\vec\alpha \cdot \vec\chi(\tau)}} \Sg^a(\tau) \:\:\:\: .
\label{ae1} \eneq In Eq.\eqref{ae1} the colons $\nor{\ldots}$
denote normal ordering with respect to the vacuum of the bosonic
theory \cite{giuso1}, with all the expectation values computed
with respect to the WFP Hamiltonian, and
$\mathbb{A}=\{\pm\vec\alpha_1,\pm\vec\alpha_2,\pm\vec\alpha_3\}$.
Furthermore, Eq.\eqref{ae1} defines the dimensionless couplings
$g_{\vec\alpha}^a$.

Expanding $\mathcal{Z}$ in the couplings $g_{\vec\alpha}^a$ yields
\begin{align}
  \Z &= \Z_{(0)} \Bigg( 1 + \frac{1}{2}\sum_{\vec\alpha \in \mathbb{A}} 
\sum_{a,b = x,y,z} L^{-2} g^a_{\vec\alpha} g^b_{-\vec\alpha}\iint d\tau d\tau'\:
  \langle \nor{e^{i\vec{\alpha}\cdot\vec{\chi}(\tau')}}\nor{e^{-i\vec{\alpha}
\cdot\vec{\chi}(\tau)}} \rangle \langle \Sg^a(\tau')\Sg^b(\tau) \rangle + \\
  &- \frac{1}{3!}\sum_{\substack{\vec\alpha,\vec\beta,\vec\gamma \in \mathbb{A} \\ \vec\alpha+\vec\beta+\vec\gamma=0}}\sum_{a,b,c}
  L^{-3} g^a_{\vec\alpha} g^b_{\vec\beta} g^c_{\vec\gamma}\iiint d\tau d\tau' d\tau''\:
  \langle \nor{e^{i\vec{\alpha}\cdot\vec{\chi}(\tau'')}}\nor{e^{i\vec{\beta}\cdot\vec{\chi}(\tau')}}
  \nor{e^{i\vec{\gamma}\cdot\vec{\chi}(\tau)}} \rangle \langle \Sg^a(\tau'')\Sg^b(\tau') \Sg^c(\tau) \rangle + \ldots
  \Bigg)\;\; ,
  \label{eq:partexp}
\end{align}
where we have used the fact that $\langle \nor{e^{i\vec{\alpha}^{(1)}\cdot\vec{\chi}(\tau')}}\ldots\nor{e^{i\vec{\alpha}^{(n)}\cdot\vec{\chi}(\tau)}} \rangle$
is different from $0$ only if $\sum_{j=1}^n\vec\alpha^{(j)}=0$.
Following the procedure of Refs.\cite{Cardy1996scaling,giuso2}, one finds
that  a rescaling of $L$ implies a renormalization of the couplings,
according to the RG flow equations that, to second order in
the boundary coupling strenghts, are determined by the short-distance
operator product expansions (O.P.E.s) 
\begin{equation}
  \left\{ \nor{ e^{ \pm i \vec{\alpha}_i \cdot \chi ( \tau ) } }
  \nor{ e^{ \pm i \vec{\alpha}_j \cdot \chi ( \tau' ) } } \right\}_{ \tau'
\to \tau^- } \approx \left[ \frac{u \lvert \tau - \tau' \rvert}{ L} \right]^{
- \frac{1}{g}} \nor{e^{ \mp i \vec{\alpha}_k \cdot \vec{\chi} ( \tau )}}
\;\;\;\; ,
\label{ae4}
\end{equation}
\noindent
when $\vec\alpha_i+\vec\alpha_j+\vec\alpha_k = 0$, and  
\beq
:e^{ \pm i \vec{\alpha}_j \cdot \vec{\chi} ( \tau )}: \: : 
e^{ \mp  i \vec{\alpha}_j \cdot \vec{\chi} ( \tau' )} : \approx_{\tau' \to \tau^-} 
{\rm const} \pm \left| \frac{u ( \tau - \tau') }{L} \right|^{ - \frac{1}{g}} \left\{  \left| \frac{u ( \tau - \tau') }{L} \right|^{ 1- \frac{1}{g}} 
\: i \vec{\alpha}_j \cdot \left[ \frac{L}{u} \frac{\partial \vec{\chi} ( \tau) }{ \partial \tau}  \right] \right\}
\:\:\:\: ,
\label{ae4.b}
\eneq
\noindent
as well as by the O.P.E.s between the spin-$1/2$ operators
which, may be derived starting from the equations of motion for a spin
in magnetic field $B=\sqrt{B_\parallel^2+B_x^2+B_y^2}$, given by:
\begin{align}
  \Sg^x(\tau) &= \Sg^x(0) \left( - 2\cos^2\phi \sin^2\theta \sinh^2\left( \frac{B\tau}{2} \right) + \cosh (B\tau) \right) + \nn\\
  &+ \Sg^y(0) \left( i\cos\theta \sinh(B\tau) - \sin(2\phi)\sin^2\theta \sinh^2\left( \frac{B\tau}{2} \right)  \right) + \nn \\
  &- \Sg^z(0) \left( i\sin\theta\sin\phi \sinh(B\tau) + \sin 2\theta \cos\phi \sinh^2\left( \frac{B\tau}{2} \right)  \right)\;\; , \\
  \Sg^y(\tau) &= \Sg^y(0) \left( - 2\sin^2\phi \sin^2\theta \sinh^2\left( \frac{B\tau}{2} \right) + \cos^2\theta\cosh(B\tau) \right) + \nn\\
  &+ \Sg^z(0) \left( i\cos\phi \sin\theta \sinh(B\tau) - \sin\phi \sin(2\theta) \sinh^2\left( \frac{B\tau}{2} \right)  \right) + \nn \\
  &- \Sg^x(0) \left( i\cos\theta \sinh(B\tau) + \sin(2\phi) \sin^2\theta \sinh^2\left( \frac{B\tau}{2} \right)  \right)\;\; , \\
  \Sg^z(\tau) &= \Sg^z(0) \left(1 + 2\sin^2\theta \sinh^2\left( \frac{B\tau}{2} \right) \right) + \nn\\
  &+ \Sg^x(0) \left( i\sin\phi \sin\theta \sinh(B\tau) - \cos\phi \sin(2\theta) \sinh^2\left( \frac{B\tau}{2} \right)  \right) + \nn \\
  &- \Sg^y(0) \left( i\cos\phi \sin\theta \sinh(B\tau) + \sin \phi \sin(2\theta) \sinh^2\left( \frac{B\tau}{2} \right)  \right) \;\; ,
\label{corref1}
\end{align}
\noindent
where the angles $\phi$ and $\theta$ are defined by
\begin{equation}
  B_\parallel = B \cos\theta \; ,\;\; B_x = B \sin\theta \cos\phi \;,\;\; B_y = B \sin\theta \sin\phi\;\;.
\end{equation}
The resulting O.P.E.s are:
\begin{subequations}
\begin{align}
  \left\{ \Sg^x(\tau') \Sg^x(\tau) \right\}_{\tau'\rightarrow\tau} &\approx
  \frac{1}{4} + \frac{B}{2}(\tau' - \tau)\left[ \Sg^z(\tau) \cos\theta +  \Sg^y (\tau) \sin\phi \sin\theta \right]\nn\\
  \left\{ \Sg^y(\tau') \Sg^y(\tau) \right\}_{\tau'\rightarrow\tau} &\approx
  \frac{1}{4} + \frac{B}{2}(\tau' - \tau)\left[ \Sg^z(\tau) \cos\theta + \Sg^x (\tau) \cos\phi \sin\theta \right]\nn\\
  \left\{ \Sg^z(\tau') \Sg^z(\tau) \right\}_{\tau'\rightarrow\tau} &\approx
  \frac{1}{4} + \frac{B}{2}(\tau' - \tau)\left[ \Sg^z(\tau) \cos\phi \sin\theta + \Sg^y (\tau) \sin\phi \sin\theta \right]\nn\\
  \left\{ \Sg^x(\tau') \Sg^y(\tau) \right\}_{\tau'\rightarrow\tau} &\approx
  \frac{i}{2} \Sg^z(\tau) + \frac{B}{2}(\tau' - \tau)\left[ \frac{i}{2} {\bf I} \cos\theta + \Sg^x (\tau) \sin\phi \sin\theta \right]\nn\\
  \left\{ \Sg^y(\tau') \Sg^z(\tau) \right\}_{\tau'\rightarrow\tau} &\approx
  \frac{i}{2} \Sg^x(\tau) + \frac{B}{2}(\tau' - \tau)\left[ \frac{i}{2} {\bf I}  \sin\theta \cos\phi + \Sg^y (\tau) \cos\theta \right]\nn\\
  \left\{ \Sg^z(\tau') \Sg^x(\tau) \right\}_{\tau'\rightarrow\tau} &\approx
  \frac{i}{2} \Sg^y(\tau) + \frac{B}{2}(\tau' - \tau)\left[ \frac{i}{2} {\bf I}  \sin\phi \sin\theta + \Sg^z (\tau) \cos\phi \sin\theta \right]\nn\\
  \left\{ \Sg^x(\tau') \Sg^z(\tau) \right\}_{\tau'\rightarrow\tau} &\approx
  -\frac{i}{2} \Sg^y(\tau) + \frac{B}{2}(\tau' - \tau)\left[ -\frac{i}{2} {\bf I}  \sin\phi \sin\theta + \Sg^x (\tau) \cos\theta \right]\nn\\
  \left\{ \Sg^y(\tau') \Sg^x(\tau) \right\}_{\tau'\rightarrow\tau} &\approx
  -\frac{i}{2} \Sg^z(\tau) + \frac{B}{2}(\tau' - \tau)\left[ -\frac{i}{2} {\bf I}  \cos\theta + \Sg^y (\tau)  \cos\phi \sin\theta \right]\nn\\
  \left\{ \Sg^z(\tau') \Sg^y(\tau) \right\}_{\tau'\rightarrow\tau} &\approx
  -\frac{i}{2} \Sg^x(\tau) + \frac{B}{2}(\tau' - \tau)\left[ -\frac{i}{2} {\bf I}  \cos\phi \sin\theta + \Sg^z (\tau) \sin\phi \sin\theta \right]
  \:\:\:\: .
\label{spinc2}
\end{align}
\noindent
When $B_\parallel = B_\perp = 0$ spin operators do not evolve in
imaginary time and Eq.(\ref{spinc2}) reduces to the usual spin
algebra
\begin{equation}
  {\bf S}_G^a {\bf S}_G^b = \frac{1}{4}\delta^{ab} {\bf I} + \frac{1}{2} i \epsilon^{abc} {\bf S}_G^c
\:\:\:\: .
\end{equation}
\label{ae5}
\end{subequations}
\noindent 
Since at the WFP the fields $\vec\chi(\tau)$ and $\vec\Sg(\tau)$
are decoupled, from Eq.(\ref{ae4},\ref{ae4.b}) 
one finds that two different  
contributions to the boundary action are generated
to the second order in the boundary coupling  strenghts; 
to evidence these contributions, one rewrites Eq.(\ref{eq:hbfermion})
as

\beq
S_{\bf B}^{(I)} = \sum_j \int_0^\beta \: d \tau \left\{ {\cal E}_j [ {\bf S}_G ( \tau ) ] 
e^{ i \vec{\alpha}_j \cdot \vec{\chi} ( \tau )} + \bar{\cal E}_j [ {\bf S}_G ( \tau ) ] 
e^{ - i \vec{\alpha}_j \cdot \vec{\chi} ( \tau )} \right\}
\:\:\:\: , 
\label{eis}
\eneq
\noindent
with  the ${\bf S}_G$-dependent 
couplings  ${\cal E}_j [ {\bf S}_G ]$ and  $\bar{\cal E}_j [ {\bf S}_G ]$ given by

\begin{eqnarray}
{\cal E}_j [ {\bf S}_G  ] &=& E_1 + 2 i E_z {\bf S}_G^z + 2 E_3 \left\{ \cos \left[ \frac{2 \pi}{3} ( j - 2 ) \right]
{\bf S}_G^x + \sin \left[ \frac{2 \pi}{3} ( j - 2 ) \right]
{\bf S}_G^y  \right\}  \nonumber \\
 \bar{\cal E}_j [ {\bf S}_G  ] &=& E_1 - 2 i E_z {\bf S}_G^z + 2 E_3 \left\{ \cos \left[ \frac{2 \pi}{3} ( j - 2 ) \right]
{\bf S}_G^x + \sin \left[ \frac{2 \pi}{3} ( j - 2 ) \right]
{\bf S}_G^y  \right\} 
\:\:\:\: . 
\label{ae5.b}
\end{eqnarray}
\noindent
 ${\cal E}_j [ {\bf S}_G ]$ and  $\bar{\cal E}_j [ {\bf S}_G ]$
contribute to the boundary action in Eq.(\ref{ae1}) through 
the  O.P.E. as

\[
 \sum_{ j , j'} \bar{\cal E}_j [ {\bf S}_G ( \tau ) ] 
: e^{ - i \vec{\alpha}_j \cdot \vec{\chi} ( \tau )} : {\cal E}_{j'} [ {\bf S}_G ( \tau' ) ] 
: e^{ - i \vec{\alpha}_{j'} \cdot \vec{\chi} ( \tau' )} :  \approx_{\tau' \to \tau^-} 
\left| \frac{u ( \tau - \tau' ) }{L} \right|^{ - \frac{1}{g} }  \sum_k \biggl\{ \biggl[ E_1^2 - 4 E_z^2 - 2 E_3^2 
\]
\beq
- 4 i E_z E_1 {\bf S}_G^z ( \tau ) - 2 E_1 E_3 \left( \cos \left[ \frac{2 \pi}{3} ( k - 2 ) \biggr]
{\bf S}_G^x ( \tau ) + \sin \biggl[ \frac{2 \pi}{3} ( k - 2 ) \biggr]
{\bf S}_G^y ( \tau ) \right) \right] : e^{  i \vec{\alpha}_k \cdot \vec{\chi} ( \tau ) } : \biggr\} 
\;\;\;\; , 
\label{ae5.c}
\eneq
\noindent
and

 \[
\sum_{j  } \{ {\cal E}_j [ {\bf S}_G ( \tau ) ] 
: e^{  i \vec{\alpha}_j \cdot \vec{\chi} ( \tau )} :  \bar{\cal E}_{j} [ {\bf S}_G ( \tau' ) ] 
: e^{ - i \vec{\alpha}_{j} \cdot \vec{\chi} ( \tau' )} :  +  \bar{\cal E}_{j} [ {\bf S}_G ( \tau  ) ] 
: e^{  - i \vec{\alpha}_j \cdot \vec{\chi} ( \tau )} :  {\cal E}_{j} [ {\bf S}_G ( \tau' ) ] 
: e^{  i \vec{\alpha}_{j} \cdot \vec{\chi}  ( \tau' )} :  \}
\approx_{\tau' \to \tau^-} 
\]
\beq
- \left| \frac{u ( \tau - \tau' ) }{L} \right|^{ 1- \frac{2}{g} } \sum_j \left\{
4 E_z E_3 \left[  \cos \left[ \frac{2 \pi}{3} ( j - 2 ) \right] {\bf S}_G^y - 
\sin \left[ \frac{2 \pi}{3} ( j - 2 ) \right] {\bf S}_G^x  \right] i \vec{\alpha}_j \cdot 
\left[ \frac{L}{u} \frac{\partial \vec{\chi} ( \tau) }{\partial \tau} \right]  \right\}
\:\:\:\:.
\label{ae5.d}
\eneq
\noindent
From Eq.(\ref{ae5.d}), one sees that, though terms proportional to  
$\frac{ \partial \vec{\chi}}{ \partial \tau}$  are not present in Eq.(\ref{eq:hbfermion}), 
they are dynamically generated by the renormalization
group procedure, yielding a new term given by   

\beq
\delta S_{\bf B}^{(I)} = i \Lambda \sum_{ j } \: \int_0^\beta \: d \tau \: \left\{ 
\cos \left[ \frac{2 \pi}{3} ( j - 2 ) \right] {\bf S}_G^y ( \tau ) - 
\sin \left[ \frac{2 \pi}{3} ( j - 2 ) \right] {\bf S}_G^x ( \tau ) 
\right\} \vec{\alpha}_j \cdot \left[ \frac{L}{u} \frac{\partial \vec{\chi} ( \tau ) }{\partial \tau} \right]
\:\:\:\: .
\label{ae5.e}
\eneq
\noindent
The boundary coupling $\Lambda$ is renormalized from the 
terms in Eq.(\ref{ae5.d}). 

For a set of (dimensionless) boundary couplings
$g_\alpha^a$, the RG equations to the second order in
the couplings  are given by:
\begin{equation}
  \frac{d g_{  \alpha}^a(l)}{dl} = \left( 1-h_{  \alpha} \right) g_{\alpha}^a(l) +
  \sum_{ \beta, \gamma }
\sum_{b,c=x,y,z} C_{\beta,\gamma,\alpha}^{b,c,a}\, g_{\beta}^b(l) g_{\gamma}^c(l)
  \label{eq:rgflow}
\end{equation}
where $l=\ln\left( L/L_0 \right)$, $L_0$ is a reference length scale,
 $h_{ \alpha}$ is the scaling dimension of the boundary operator associated to the coupling
$g_ \alpha^a$ and the coefficient  $C_{  \beta, \gamma,  \alpha}^{b,c,a}$ 
are defined by  the   O.P.E.s. Thus, 
deriving the coefficients
$C_{\beta,\gamma,\alpha}^{b,c,a}$  from Eqs.(\ref{ae5.c},\ref{ae5.d})
and taking into account that $h= \frac{1}{g}$ ($h=1$) for any vertex operator
$e^{ \pm i \vec{\alpha}_j \cdot \vec{\chi} ( \tau )}$ (for 
$\frac{\partial \chi_1 ( \tau)}{ \partial \tau} , \frac{\partial \chi_2  ( \tau)}{ \partial \tau}$),
one finds that the RG equations for the
dimensionless couplings $G_1 ( L ) = L^{1 - \frac{1}{g}} E_1$,
$G_z ( L ) = L^{1 - \frac{1}{g}} E_z$,  $G_3 ( L ) = L^{1 -
\frac{1}{g}} E_3$, $G_\Lambda ( L ) = \Lambda$,   are given by

\begin{align}
 \frac{d G_1 ( l) }{d l} &=
\left[ 1 - \frac{1}{g} \right] G_1 ( l)
- 2 \left[ G_1^2 ( l ) - G_z^2 ( l ) - \frac{G_3^2 ( l)}{2}\right]
\equiv \beta_1 ( G_1 , G_z , G_3 , G_\Lambda))\nn \\
  \frac{d G_z ( l ) }{d l}&=
\left[ 1 - \frac{1}{g} \right] G_z ( l )
 + 4 G_1 ( l ) G_z ( l ) + 2 G_3 ( l ) G_\Lambda ( l ) 
\equiv \beta_z ( G_1 , G_z , G_3 , G_\Lambda )\nn \\
 \frac{d G_3 ( l) }{d l}&=
\left[ 1 - \frac{1}{g} \right] G_3 ( l) + 2
G_1 ( l ) G_3 ( l) + 2 G_z ( l ) G_\Lambda ( l ) 
\equiv \beta_3 ( G_1 , G_z , G_3 , G_\Lambda )\nn \\
 \frac{d G_\Lambda ( l) }{d l}&= 4 G_z ( l ) G_3 ( l) \equiv \beta_\Lambda ( G_1 , G_z , G_3 , G_\Lambda )
\:\: .
\label{ae7}
\end{align}
\noindent
From Eqs.(\ref{ae7}) one sees that no linear term appears in the $\beta$-function
for the running coupling $G_\Lambda$, since the
corresponding boundary operator has scaling dimension $ 1$, while the ones containing
the vertex operators have scaling dimension $  1 / g$. Furthemore, looking at the linear term 
in the $\beta$-functions in Eqs.(\ref{ae7}), one sees that 
the couplings $G_1 ( l ) , G_z ( l ) , G_3 ( l )$ are all
irrelevant for $g<1$ and that  $G_\Lambda ( l )$ scales roughly  like
$e^{ ( 2 - \frac{2}{g} ) l}$, so that, for $g<1$, the term $\propto \frac{\partial \vec{\chi} ( \tau )}{ \partial \tau}$
is irrelevant, as well. At variance, for $g>1$,  $G_1 ( l ) , G_z ( l ) , G_3 ( l )$ 
are all relevant, while $G_\Lambda ( l )$ is only marginally relevant. As a result, 
$\delta S_{\bf B}^{(I)}$ is subleading with respect to $S_{\bf B}^{(I)}$, and,
for $g>1$,  one may write Eqs.(\ref{ae7})  by keeping
only leading  contributions to the boundary interaction; thus one gets
 
\begin{align}
 \frac{d G_1 ( l) }{d l} &=
\left[ 1 - \frac{1}{g} \right] G_1 ( l)
- 2 \left[ G_1^2 ( l ) - G_z^2 ( l ) - \frac{G_3^2 ( l)}{2}\right]
\equiv \beta_1 ( G_1 , G_z , G_3 )\nn \\
  \frac{d G_z ( l ) }{d l}&=
\left[ 1 - \frac{1}{g} \right] G_z ( l )
 + 4 G_1 ( l ) G_z ( l )
\equiv \beta_z ( G_1 , G_z , G_3 )\nn \\
 \frac{d G_3 ( l) }{d l}&=
\left[ 1 - \frac{1}{g} \right] G_3 ( l) + 2
G_1 ( l ) G_3 ( l)
\equiv \beta_3 ( G_1 , G_z , G_3 )
\:\: .
\label{ae7.b}
\end{align}
\noindent
From Eqs.(\ref{ae7.b}), one sees that, when $g>1$, the boundary interaction provides a
relevant perturbation and that the pertubative approach breaks down when the
leads reach the healing length $L_*$, for which the biggest
dimensionless coupling is of order $L_0^{1 - \frac{1}{g}} E_J$.
$L_0$ is the reference length such that  $G_i(l = 0 )=L_0^{1 -
\frac{1}{g}}  E_i$, with the $E_i$'s
defined in Eq.\eqref{eq:couplings}; since Eq.\eqref{eq:couplings}
shows  that the biggest coupling is $G_3(l)$, 
$L_{*}$ should be determined from the requirement that $G_3
\left[  \ln \left( \frac{L_*}{L_0} \right) \right] \sim L_0^{1 - \frac{1}{g}} E_{J}$ yielding
\begin{equation}
  L_* \sim \left( \frac{E_{J}}{E_3} \right)^{\frac{g}{g-1}} L_0 \;\; .
  \label{eq:lstar}
\end{equation}
\noindent

It should be noticed that, for $g=1$,  one should use  
Eqs.(\ref{ae7}), instead of Eqs.(\ref{ae7.b}), as all the four couplings 
$G_1 ( l ) , G_z ( l ) , G_3 ( l ) , G_\Lambda ( l )$ correspond to
marginally relevant perturbations. To spell out the RG flows  for $g=1$,
one should notice that $G_z (L_0 )=0$ if $H=0$; from 
 Eqs.(\ref{ae7}), then, the RG flows run  along the
manifold in parameter space defined by $G_z (l ) = G_\Lambda ( l) = 0$
and by $G_3 ( l ) = 2 G_1 ( l) $. Since $G_3$ depends on $l$ as

\beq
\frac{d G_3 ( l) }{d l} = [ G_3 ( l) ]^2
\;\; ,
  \label{eq:lstar.2}
\end{equation}
\noindent
one sees that  the RG equations in
Eq.(\ref{ae7}) coincide with the ones obtained for the boundary
coupling in the isotropic Kondo model. Thus,
for $g=1$ and $G_z  ( L_0 ) =0$, the tetrahedral JJN 
simulates a Kondo spin with isotropic couplings to the
electrons and $L_*$ coincides with the Kondo length $L_K$, 
given by $L_K \sim L_0 e^{ \frac{1}{G_3 ( L_0 )}}$. A
 small value of $H$ should just slightly change
the RG flow, possibly to the one corresponding to a 
Kondo spin with anisotropic couplings to the band electrons
\cite{anisotropic}.

In the next section, starting from Eqs.(\ref{ae7}) we characterize
the SFP and investigate its stability against the leading boundary
perturbation.

\section{The Strong Coupled Fixed Point}
\label{strong0}

From Eq.(\ref{ae7}) one sees that all the boundary couplings
become relevant when $g>1$; this implies \eqref{eq:lstar} that, as
soon as  $L \geq L_*$ ,
the running boundary couplings cross over towards strong coupling.
At the SFP, the plasmon fields $\vec{\chi} ( 0 )$ satisfy
Dirichlet boundary conditions since  their values must
coincide with a minimum of ${\mathcal H}_{\bf B}$; in addition,
the leading boundary interaction is a combination of phase slip
(instanton) operators describing tunnelling events between the
neighboring minima of  ${\mathcal H}_{\bf B}$.

\subsection{Minima of ${\mathcal H}_{\bf B}$}
\label{strong1}

Dirichlet boundary conditions are set  by requiring that $\chi_1 (
0 ) , \chi_2 ( 0 )$ take values corresponding to a minimum of
${\mathcal H}_{\bf B}$. To determine the set of minima of
${\mathcal H}_{\bf B}$ it is most convenient to represent
\cite{chamon,chamon2} the spin-$1/2$ operator ${\bf S}_G$ as a
2$\times$2 matrix and rewrite ${\mathcal H}_{\bf B}$ as

\begin{equation}
  \hb = \begin{pmatrix}
    V_1(\vec{\chi} (0) )+V_z(\vec{\chi} (0)) + B_{\parallel}(f) && V_x(\vec{\chi} (0)
)-i V_y(\vec{\chi} (0)) \\
    V_x(\vec{\chi} (0))+i V_y(\vec{\chi} (0) ) &&
    V_1(\vec{\chi} (0) )-V_z(\vec{\chi} (0)) - B_{\parallel}(f)
  \end{pmatrix}
\:\:\:\: .
  \label{eq:hbm}
\end{equation}
In Eq.(\ref{eq:hbm}) $B_\parallel(f)$ is given by
Eq.\eqref{parallel} while, by comparison with Eq.(\ref{eq:hb2}),
the $V_i ( \vec{\chi} ( 0 ) ) $'s are determined as

\begin{alignat}{3}
 V_1 [ \vec{\chi} ] &= 2 \bar{E}_1 \sum_j \cos [ \vec{\alpha}_j \cdot \vec{\chi} ]\nn\\
 V_x [ \vec{\chi} ] &=  2 \bar{E}_3 \sum_j \cos \left[ \frac{2 \pi}{3} (j-2) \right] \cos [ \vec{\alpha}_j \cdot \vec{\chi} ] \nonumber \\
V_y [ \vec{\chi} ] &=  2 \bar{E}_3 \sum_j \sin \left[ \frac{2 \pi}{3} (j-2) \right] \cos [ \vec{\alpha}_j \cdot \vec{\chi} ]\nn\\
V_z [ \vec{\chi} ] &= - 2 \bar{E}_z  \sum_j \sin [ \vec{\alpha}_j \cdot \vec{\chi} ]
\:\:\:\: .
\label{components}
\end{alignat}
\noindent One may now easily look for the eigenvalues of $\hb$ as
a function of $\chi_1 ( 0 ) $ and $\chi_2 ( 0 ) $ obtaining
\begin{align}
  \Lambda_1(\vec{\chi} ( 0 ) )&=V_1 - \sqrt{V_x^2+V_y^2+(V_z+B_{\parallel})^2} \\
  \Lambda_2(\vec{\chi} ( 0 ) )&=V_1 + \sqrt{V_x^2+V_y^2+(V_z+B_{\parallel})^2}
\:\:\:\: .
  \label{eq:eighb}
\end{align}
\begin{figure}[htb]
\begin{center}
    \includegraphics[width=0.5\textwidth]{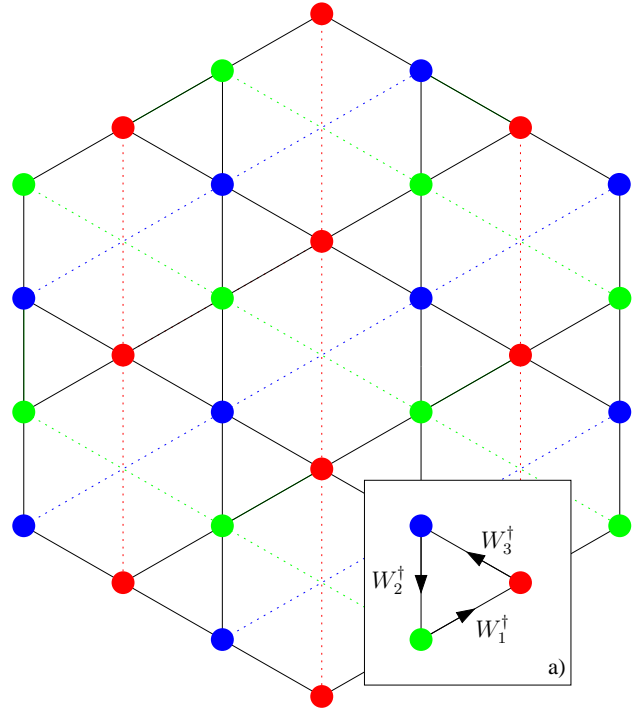}
\end{center}
  \caption{Lattice of the minima of ${\mathcal H}_{\bf B}$. In the box: the triangle whose
  vertices may be made degenerate with a pertinent choice of the external phases, and the instanton trajectories
  associated to the operators $W^\dagger_i$.}
  \label{fig:kagome}
\end{figure}
Since, for any value of $\vec{\chi}$, one has that $\Lambda_1 \leq
\Lambda_2 $ the minima of $\hb$ are obtained from the minima of
$\Lambda_1(\vec{\chi} ( 0 ) )$ only. We find that, in the $\chi_1
( 0 ) , \chi_2 ( 0 ) $-plane, these minima are located at the
vertices of the Kagome lattice displayed in Fig.\ref{fig:kagome}.
The polarization of ${\bf S}_G$ at each minimum is determined by
minimizing the boundary energy after the $V_i$'s have been
evaluated on the pertinent vertex of the Kagome lattice of minima.

 Since at the outer boundary
($x=L$) $\chi_1, \chi_2$ are connected to three bulk
superconductors at fixed phases, the corresponding boundary
conditions are given by $\chi_1 ( L , \tau ) = \mu_1 , \chi_2 ( L
, \tau ) = \mu_2$, with the relative phases $\mu_1 , \mu_2$ being
defined as $\mu_1 = ( \varphi_1 - \varphi_2 ) / \sqrt{2}$, and
$\mu_2 = ( \varphi_1 + \varphi_2 - 2 \varphi_3 ) / \sqrt{6}$. Upon
imposing Dirichlet boundary conditions for $\vec{\chi} ( x , \tau
)$ at both boundaries the mode expansion of the plasmon fields is
given by
\begin{equation}
\chi_j (x,\tau) = \xi_j + \sqrt{\frac{2}{g}}\left\{(L-x)\frac{\pi}{L}P_j -
\sum_{n\neq 0}\sin\left(\frac{\pi n x}{L}\right)\frac{\alpha_n^j}{n}e^{-  \frac{\pi}{L}n u \tau}\right\}
\:\:\:\: ,
\label{eq:mode-exp-strong}
\end{equation}
where $\xi_j=\chi_j(L)$ and $P_j$s are the zero mode operators.

The eigenvalues of $P_j$ are fixed by the boundary conditions
since $\sqrt{\frac{2}{g}}\pi P_j= \chi_j(0)-\xi_j$; they are given
by
\begin{equation}
(p_1,p_2)_l = \sqrt{2 g}\left( -\frac{\mu_1}{2 \pi} + n_{12} + \epsilon_l ,
-\frac{\mu_2}{2 \pi} + \frac{1}{\sqrt{3}} (2 n_{13} - n_{12} + \delta_l)\right)
\label{eq:zero-mode}
\;\;\;\; ,
\end{equation}
 $n_{12} , n_{13}$ being relative integers.
The index $l$ accounts for the ``red'' (R), the ``green'' (G), and
the ``blue'' (B) sublattices in Fig.\ref{fig:kagome}; $\epsilon_R
= 1$, $\epsilon_G = 1/2$, $\epsilon_B = 1/2$; $\delta_R = 0$,
$\delta_G = -1/2$, $\delta_B = 1/2$, while the constants $\xi_1 ,
\xi_2$ cancel the terms $\propto \mu_1 , \mu_2$ in $\chi_1 ( 0 ) ,
\chi_2 ( 0 )$, respectively. The spin polarization of ${\bf S}_G$
is uniform throughout each one of the three sublattices; thus, to
the minima of $\hb$ are associated the states

\begin{subequations}
\begin{align}
\left\lvert R \right\rangle &= \left( \cos\frac{\theta_f}{2} \left\lvert \Uparrow
\right\rangle + e^{i\frac{\pi}{3}} \sin\frac{\theta_f}{2} \left\lvert \Downarrow \right\rangle \right)\,, \\
  \left\lvert  G \right\rangle &= \left( \cos\frac{\theta_f}{2} \left\lvert \Uparrow \right\rangle
  + e^{-i\frac{\pi}{3}} \sin\frac{\theta_f}{2} \left\lvert \Downarrow \right\rangle \right)\,, \\
  \left\lvert B \right\rangle &= \left( \cos\frac{\theta_f}{2} \left\lvert \Uparrow \right\rangle
  + e^{i\pi} \sin\frac{\theta_f}{2} \left\lvert \Downarrow \right\rangle \right)\,,
\end{align}
  \label{eq:eigstathb}
\end{subequations}
where
$\cos\frac{\theta_f}{2}=\frac{\sqrt{16+B_{\parallel}^2}-B_{\parallel}}{\sqrt{16+B_{\parallel}^2-B_{\parallel}\sqrt{16+B_{\parallel}^2}}}$
and
$\sin\frac{\theta_f}{2}=\frac{1}{\sqrt{16+B_{\parallel}^2-B_{\parallel}\sqrt{16+B_{\parallel}^2}}}$.
To the first order in $B_{\parallel}$ (i.e. to the first order in
$(f-\pi)$) one has:
\begin{equation}
  \theta_f \approx \frac{\pi}{2} + \frac{B_{\parallel}}{4}\,.
  \label{eq:thetaf}
\end{equation}
From the zero-mode spectrum, one sees that the energy of each
field configuration gets a zero-mode contribution which is
quadratic in $\vec{p}$; inserting the solution of
Eq.(\ref{eq:mode-exp-strong}) in the noninteracting Hamiltonian
$\h_{\rm LL}$, one finds the zero-mode contribution to the total
energy for each of the three sublattices. Namely, one gets:
\begin{itemize}
 \item {\bf Red-points:}

Zero-mode eigenvalues:
\begin{equation}
(p_1 , p_2 )_R = \sqrt{2g} \left( - \frac{\mu_1}{2 \pi} + n_{12}
+1 ,  - \frac{\mu_2}{2 \pi} + \frac{1}{\sqrt{3}} ( 2n_{13} - n_{12} ) \right)
\;\;\;\; ,
\label{cm1}
\end{equation}
\noindent
Zero-mode contribution to the total energy:

\begin{equation}
E^{(0)}_{R , n_{12} , n_{13} } ( \vec{\mu} ) = \frac{\pi u g}{L}
\left[  \left( - \frac{\mu_1}{2 \pi} + n_{12}
+1 \right)^2 + \left( - \frac{\mu_2}{2 \pi} +
\frac{1}{\sqrt{3}} ( 2n_{13} - n_{12} ) \right)^2 \right]
\:\:\:\: .
\label{cm2}
\end{equation}
\noindent

 \item {\bf Green-points:}

Zero-mode eigenvalues:

\begin{equation}
(p_1 , p_2 )_G = \sqrt{2g} \left( - \frac{\mu_1}{2 \pi} + n_{12}
+ \frac{1}{2},  - \frac{\mu_2}{2 \pi} + \frac{1}{\sqrt{3}} ( 2n_{13} - n_{12} - \frac{1}{2}) \right)
\;\;\;\; ,
\label{cm3}
\end{equation}
\noindent

Zero-mode contribution to the total energy:
\begin{equation}
E_{G , n_{12} , n_{13} } ( \vec{\mu} ) = \frac{\pi u g}{L}
\left[  \left( - \frac{\mu_1}{2 \pi} + n_{12}
+\frac{1}{2}  \right)^2 + \left( - \frac{\mu_2}{2 \pi} +
\frac{1}{\sqrt{3}} ( 2n_{13} - n_{12} - \frac{1}{2}) \right)^2 \right]
\:\:\:\: .
\label{cm4}
\end{equation}
\noindent

\item {\bf Blue-points:}

Zero-mode eigenvalues:

\begin{equation}
(p_1 , p_2 )_B = \sqrt{2g} \left( - \frac{\mu_1}{2 \pi} + n_{12}
+ \frac{1}{2},  - \frac{\mu_2}{2 \pi} + \frac{1}{\sqrt{3}} ( 2n_{13} - n_{12} + \frac{1}{2}) \right)
\;\;\;\; ,
\label{cm5}
\end{equation}
\noindent

Zero-mode contribution to the total energy:
\begin{equation}
E_{B , n_{12} , n_{13} } ( \vec{\mu} ) = \frac{\pi u g}{L}
\left[  \left( - \frac{\mu_1}{2 \pi} + n_{12}
+\frac{1}{2}  \right)^2 + \left( - \frac{\mu_2}{2 \pi} +
\frac{1}{\sqrt{3}} ( 2n_{13} - n_{12} + \frac{1}{2}) \right)^2 \right]
\:\:\:\: .
\label{cm6}
\end{equation}
\noindent
\end{itemize}
Through the external phases $\mu_1 , \mu_2$ one adds an effective
quadratic potential breaking the degeneracy of the zero-modes. It
is quite remarkable that the zero-mode energy spectrum reported in
Eqs.(\ref{cm2},\ref{cm4},\ref{cm6}) displays, for convenient
choices of the applied phases $\mu_1 , \mu_2$, a threefold
degeneracy between the three sites lying at the vertices of a
single triangle (box in Fig.\ref{fig:kagome}). As we shall see in
later sections, this feature of a tetrahedral JJN is crucial for
its applications to quantum information processing tasks.

\subsection{Leading boundary perturbation at the SFP} \label{strong2}

In order to describe effects of the leading boundary perturbation
at the SFP one should rewrite the boundary Hamiltonian in terms of
the phase-slip operators representing instanton trajectories
connecting nearest neighboring sites on the lattice in
Fig.\ref{fig:kagome}. For this purpose, it is most convenient to
introduce the phase slip field operators $\Theta_1 ( x , t ) ,
\Theta_2 ( x , t)$, dual to $\chi_1 ( x , t ) , \chi_2 ( x , t )$,
and, thus, defined  by \cite{giuso4}

\begin{equation}
\frac{1}{u g } \frac{\partial \Theta_i ( x , t ) }{ \partial t}  =
\frac{\partial \chi_i ( x , t )}{ \partial x} \;\;\; , \;\;
\frac{g}{ u } \frac{\partial \chi_i ( x , t ) }{ \partial t}  =
\frac{\partial \Theta_i ( x , t )}{ \partial x}
\:\:\:\: .
\label{dual1}
\end{equation}
\noindent In particular, Eqs.(\ref{dual1}) imply that
$\Theta_{1,2} ( x , t )$ obey Neumann boundary conditions at $x=0$
and at $x=L$. As a result, the mode expansion of the instanton
fields is given by

\begin{equation}
  \Theta_j ( x , t ) = \sqrt{2g} \left\{ \theta_0^j + \frac{  \pi u t }{L} P_j
+ i \sum_{ n \neq 0} \cos \left[ \frac{ \pi n x}{L} \right]
\frac{ \tilde{\alpha}_n^j}{n}  e^{ - i \frac{ \pi }{L} n v t } \right\}
  \label{eq:dual-mode-exp}
\:\:\:\: ,
\end{equation}
\noindent with

\begin{equation}
[ P_i , \theta_0^j ] = - i \delta_{i , j} \;\;\; , \;\;
[ \tilde{\alpha}_n^i , \tilde{\alpha}_m^j ] = n \delta_{n+m , 0 } \delta_{i , j }
\:\:\:\: .
\label{algdual}
\end{equation}
\noindent A quantum tunnelling between two adjacent minima lying
on the lattice shown in Fig.\ref{fig:kagome} involves a quantum
jump between different zero mode eigenstates. The quantum phase
slip operators corresponding to the allowed quantum jumps, $W_i$,
$W_i^\dagger$ ($i=1,2,3$),  are given by:
\begin{align}
  W_1^\dag &=\nor{\exp\left[\frac{i}{\sqrt{3}}\vec{\rho}_1\cdot\vec{\Theta}(0)\right]} \nn \\
W_2^\dag &=\nor{\exp\left[\frac{i}{\sqrt{3}}\vec{\rho}_2\cdot\vec{\Theta}(0)\right]} \nn \\
W_3^\dag
&=\nor{\exp\left[\frac{i}{\sqrt{3}}\vec{\rho}_3\cdot\vec{\Theta}(0)\right]}.
  \label{eq:istantons}
\;\;\;\; ,
\end{align}
The vectors  $\vec{\rho_i}$ are defined so that
$\sqrt{2g}\frac{\vec{\rho_i}}{\sqrt{3}}$ is the ``distance''
between nearest neighboring eigenvalues of $\vec{P}$, as
determined by the commutation relations

\begin{equation}
[ \vec{P} , W_i ] = - \sqrt{\frac{2g}{3}} \vec{\rho}_i W_i\;\;\; ,\;\;
[ \vec{P} , W_i^\dagger ] =  \sqrt{\frac{2g}{3}} \vec{\rho}_i W_i^\dagger
\:\:\:\: ,
\label{dualcom}
\end{equation}
which yield

\begin{equation}
\vec{\rho}_1 = \left( \begin{array}{c} \sqrt{3}/2 \\  1/2 \end{array} \right)
\;\;\; , \;\;
\vec{\rho}_2 = \left( \begin{array}{c} 0  \\ -1 \end{array} \right)
\;\;\; , \;\;
\vec{\rho}_3 = \left( \begin{array}{c} -\sqrt{3}/2 \\ 1/2 \end{array} \right)
\:\:\:\: .
\label{dualvec}
\end{equation}
\noindent In Fig.\ref{fig:kagome} we have represented the quantum
jumps between  the eigenvales of $\vec{P}$ corresponding to the
operators $W_i^\dag$; the action of the hermitean conjugate
operators  $W_i$ may be simply represented by reversing the
arrows.

Since a tunnelling event between minima of ${\mathcal H}_{\bf B}$
involves a rotation in the two dimensional space spanned by the
eigenstates of ${\bf S}_G$, and since the allowed directions of
tunnelling from a given minimum depend on the position of this
minimum in the Kagome lattice, each quantum phase slip operator
has to be multiplied by the spin operator mapping the state
$\{\left\lvert i\right\rangle \}$ ($i=R,G,B,$) onto the other two
states. As a result the leading boundary perturbation at the SFP
is given by

\begin{equation}
  \tilde{\mathcal H}_{\bf B} = - Y e^{ i \gamma} \{  S_{GR}^\dagger  W_{1}^{\dagger}   +
S_{BG}^\dagger    W_2^{\dagger}  + S_{BR}^\dagger
   W_3^{\dagger}  \}  +  {\rm h.c.}
\:\:\:\: , \label{stronx2}
 \end{equation}
\noindent with $S_{ij}$ the operator sending the ``j`` spin state
into the ``i`` one,  while the parameters $Y , \gamma$ are computed in
\ref{appe3}. Since the scaling dimension of the operators $ \{
W_i, W_i^\dag \}$ is $\frac{g}{3}$ the running coupling strength
for the dual boundary coupling may be defined as $\zeta ( L ) = y
(L) e^{ i \gamma}$, with $y(L)=L Y(L)$.

Even if $\zeta$ is, in general, a complex coupling strength, the
renormalization group equations may be derived following the
standard procedure used in \cite{giuso4} which, starting from
Eq.(\ref{stronx2}), allows to determine the euclidean dual
boundary action at the SFP as

\begin{equation}
\tilde{S}_{\bf B} = - Y e^{ i \gamma} \int_0^\beta \: d \tau \: \{
S_{GR}^\dagger  (\tau)  W_{1}^{\dagger}   +
S_{BG}^\dagger  ( \tau )    W_2^{\dagger}  + S_{BR}^\dagger ( \tau )
   W_3^{\dagger}  \}  +  {\rm h.c.}. \:\:\:\: ,
\label{stronxx1}
\end{equation}
\noindent Here $W_j ( \tau ) = e^{ - \frac{i}{\sqrt{3}}
\vec{\rho}_i \cdot \vec{\Theta} ( \tau ) }$, with $\Theta ( \tau )
= \Theta ( 0 , i \tau )$; one may then compute the
$\beta$-functions for the boundary coupling strengths
\cite{cardy0,giuso4} by resorting to the O.P.E.s

\begin{equation}
[ W_{1}^\dag  S^\dag ] ( \tau ) [  W_2^\dag S^\dag ] ( \tau^{'} )  \approx_{\tau^{'} \to \tau}
\left|  \frac{u ( \tau - \tau^{'} )}{L}  \right|^{-\frac{2g}{3}} [ W_{3}  S ] ( \tau )
\:\:\:\: ,
\label{stronxx2}
\end{equation}
\noindent plus cyclic permutations.

From Eq.(\ref{stronxx2}), the second-order renormalization group
equation for the running coupling $\zeta ( L )$ may be written as
\begin{equation}
\frac{d \zeta ( L ) }{d \ln \left( \frac{L}{L_0} \right)} = \left( 1 - \frac{g}{3} \right)
\zeta ( L ) - 2 e^{ - 2 i \gamma} \zeta^2 ( L )
\;\;\;\; ,
\label{stronxx3}
\end{equation}
\noindent where $L_0$ is, again, a reference length scale.
Eq.(\ref{stronxx3}) is equivalent to the system of real
differential equations for the real parameters $ y ( L ) , \gamma
( L )$ given by
\begin{align}
 \frac{d y ( L ) }{d \ln \left( \frac{L}{L_0} \right)} &= \left( 1 - \frac{g}{3} \right)
y ( L ) - 2 \cos ( 3 \gamma)  y^2 ( L ) \nonumber \\
  y(L)\frac{d \gamma ( L ) }{d \ln \left( \frac{L}{L_0} \right)} &= 2 \sin ( 3 \gamma ( L ))
y^2 ( L ) \;\;\;\; .
\label{finaleqx}
\end{align}
\noindent From Eqs.(\ref{finaleqx}) one sees that the phase
$\gamma$ is renormalized only to the second order in the boundary
couplings and that, for $\gamma = k \pi /3$ ($k$ integer),  there
are lines of fixed points in the $y-\gamma$-plane; furthermore,
the line $\gamma=\frac{\pi}{3}$ is made of attractive fixed
points.

\begin{figure}
\begin{center}
    \includegraphics[width=0.55\textwidth]{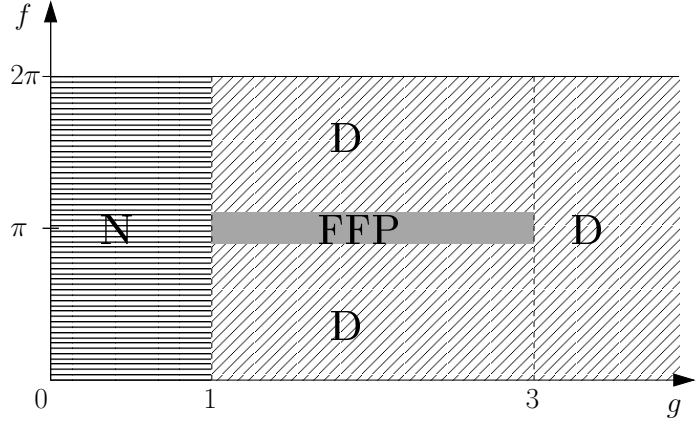}
\end{center}
  \caption{Phase diagram in the strip in the $g , f$-plane corresponding to
$0 \leq f \leq 2 \pi$: the weakly-coupled phase (corresponding to
Neumann-(N)- boundary conditions at the inner boundary) is stable
for $g<1$ and for any $f$; the strongly-coupled phase
(corresponding to Dirichlet-(D)- boundary conditions at the inner
boundary) is stable for $g>1$ and for any $f$. For $g<3$ and
$-\frac{\pi}{10} \lesssim (f-\pi) \lesssim \frac{\pi}{10}$ a novel
phase opens, corresponding to a stable finite coupling fixed point
(FFP). The phase diagram for $f \notin [ 0 , 2 \pi ]$ is obtained
by periodical extension of this picture.}
  \label{fig:phasediag}
\end{figure}

The phase diagram accessible to the tetrahedral JJN may be
inferred from the RG equations near the WFP, derived in section
\ref{weak0}, and the ones near the SFP, Eqs.(\ref{finaleqx}). As
displayed in Fig.\ref{fig:phasediag}, for $1 < g < 3$ and
$-\frac{\pi}{10} \lesssim (f-\pi) \lesssim \frac{\pi}{10}$
\footnote{This interval is determined in Sec.\ref{engineer0} and
accounts for the fact that, for $f\ne\pi$, the renormalization of
$B_\perp$ may lift the degeneracy between the minima.}, neither
the WFP, or the SFP, is infrared stable. As a consequence, in this
window of values of $f$ and $g$, the infrared behavior of the JJN
will be driven by an emerging FFP, whose properties will be
analyzed in the next section.

\section{The quantum doublet at the FFP}
\label{emerge0}

In this section we show that the renormalization of the instanton
tunnelling strength $Y$, due to the interaction with the plasmon
modes of the TLL leads, enforces the emergence of a quantum
doublet robust not only against the noise in the external control
parameters but also against the decoherence induced by the
coupling with the plasmon modes in the leads.

\subsection{The emerging doublet at the FFP}
\label{emerge1}

Since a real device has a finite size ($L$) the infinite
degeneracy induced at the SFP by the eigenvalues of the zero-mode
operators, is removed by finite-size effects; i.e., by the
zero-mode contributions to the total energy. As a function of the
external phases $\vec{\mu}$, the zero-mode energy associated to
each eigenvalue $\vec{p}$ is given by

\begin{equation}
E^{(0)}_{\{ R , G , B \} , n_{12} , n_{13} } [ \vec{\mu} ] =
\frac{\pi u}{2 L } [ \vec{p} ( n_{12} , n_{13} ) ]^2
\:\:\:\: .
\label{effe1}
\end{equation}
\noindent
From Eqs.(\ref{cm2},\ref{cm4},\ref{cm6}), one sees that, for a
pertinent choice of the phases $\vec{\mu}$, the three zero-mode
energies associated to the vertices of a triangle as the one drawn
in the box of  Fig.\ref{fig:kagome}, may be made degenerate. To
explicitly show this, let us set $\mu_1^* = \frac{4 \pi}{3} ,
\mu_2^* = 0$. For such a choice of the external phases, one gets
$E^{(0)}_{R  , 0 , 0  } [ \vec{\mu}^* ] = E^{(0)}_{G  , 0 , 0 } [
\vec{\mu}^* ] = E^{(0)}_{B  , 0 , 0 } [ \vec{\mu}^* ]$. Thus, if
one restricts himself to the three-dimensional subspace ${\mathcal
F}$ of the Hilbert space spanned by the states $ | R \rangle , | G
\rangle , | B \rangle$, an effective low-energy description of
the tetrahedral JJN  at the SFP may be provided by the 3$\times$3
Hamiltonian matrix $\h_{\mathcal{F}}$ given by

\begin{equation}
\h_{\mathcal{F}} = \frac{\pi g u}{9 L}  {\bf I} +
\begin{bmatrix} 0 &
                           - Y e^{ i \gamma}  & - Y e^{ - i \gamma}  \\ - Y e^{  -  i \gamma}
 & 0 & - Y e^{  i \gamma}  \\ - Y e^{  i \gamma} &
- Y e^{ - i \gamma} &  0
                          \end{bmatrix}
\:\:\:\: ,
\label{cmx1}
\end{equation}
\noindent

Form \eqref{cmx1} one sees  that, when $Y\neq 0$, the degeneracy
between $\left\lvert R \right\rangle$, $\left\lvert G
\right\rangle$ and $\left\lvert B \right\rangle$ is broken and
that the spectrum of the low-lying energy states admits as its
groundstate a quantum doublet confined away from a singlet state.
To see this, let us consider the two operators acting on
${\mathcal F}$ whose matrix representations are given by

\begin{equation}
A_1 = \begin{bmatrix} 0 & 1 & 1 \\ 1 & 0 & 1 \\ 1 & 1 & 0
             \end{bmatrix} \;\;\; , \;\;
A_2 = \begin{bmatrix} 1 & 0 & 0 \\
                                0 & 0 & e^{- i \frac{2}{3} \pi } \\
                                0 & e^{ i \frac{2}{3} \pi } & 0
\end{bmatrix}
\:\:\:\: .
\end{equation}
\noindent While  $A_1$ corresponds to a cyclic shift of the states
$ | R \rangle , | G \rangle , | B \rangle$, $A_2$ realizes the
mirror inversion about the triangle height passing through the
vertex corresponding to the $ | R \rangle$-state.  Though, as
expected,  when $ \gamma = \pi/3 $, both $A_1$ and $A_2$ commute
with $\h_{\mathcal{F}}$, $ [ \h_{\mathcal{F}} , A_1 ]
= [ \h_{\mathcal{F}} , A_2 ] = 0$, they do not commute with
each other since $[ A_1 , A_2 ] \neq 0$. This is enough to ensure
that the spectrum of $\h_{\mathcal{F}}$ must contain at least
one degenerate eigenvalue. To explicitly check it out, let us
consider the eigenvalue equation associated to $\h_{\mathcal{F}}$. It reads

\begin{equation}
- E^3 + 3 Y^2 E - 2 \cos ( 3 \gamma ) Y^3 = 0
\:\:\:\: .
\label{cmx2}
\end{equation}
\noindent For $\gamma = \pi / 3$ there is a twofold degenerate
eigenvalue $E = - Y$, and a non degenerate eigenvalue $E = 2 Y$.
The corresponding eigenstates are
\begin{align}
 | - Y \rangle_1 &= \frac{1}{\sqrt{3}} [ | R \rangle + | G \rangle + | B \rangle ]
\nonumber \\
| - Y \rangle_2 &= \frac{1}{\sqrt{3}} \left[ | R \rangle + e^{ - i \frac{2}{3} \pi} | G \rangle +
e^{ i \frac{2}{3} \pi} | B \rangle \right] \nonumber \\
| 2 Y \rangle &= \frac{1}{\sqrt{3}} \left[ | R \rangle + e^{ i \frac{2}{3} \pi} | G \rangle +
e^{ - i \frac{2}{3} \pi} | B \rangle \right]
\:\:\:\: .
\label{eigen1}
\end{align}
\noindent To rewrite $\h_{\mathcal{F}}$ using the states given in
Eq.(\ref{eigen1}), one has to rotate it to $\bar{H}_{\mathcal{F}}
= P^\dagger \h_{\mathcal{F}}P$, with the matrix $P$ given by

\begin{equation}
P = \frac{1}{\sqrt{3}}\begin{bmatrix}
            1 & 1 & 1 \\
            1 & e^{ - i \frac{2}{3} \pi} & e^{ i \frac{2}{3} \pi} \\
            1 & e^{ i \frac{2}{3} \pi} & e^{ - i \frac{2}{3} \pi}
           \end{bmatrix}
\:\:\:\: .
\label{eigen2}
\end{equation}
\noindent One obtains

\begin{equation}
P^\dagger \h_{\mathcal{F}} P =  \frac{\pi g u}{9 L}  \frac{1}{9} \mathbf{I}
+ \begin{bmatrix}
  - Y & 0 & 0 \\
  0& - Y &  0 \\
  0  &  0 & 2 Y
\end{bmatrix} \:\:\:\: .
\label{eigen3-2}
\end{equation}
\noindent Similarly, by rotating $A_2$, one gets

\begin{equation}
P^\dagger A_2 P = \begin{bmatrix} 0 & 1 & 0 \\ 1 & 0 & 0 \\ 0 & 0 & 1
                         \end{bmatrix}
\;\;\;\; ;
\label{eigenx1}
\end{equation}
\noindent thus, $A_2$ is the operator swapping the degenerate
states $ | - Y \rangle_1 ,  | - Y \rangle_2 $ with each other.

Eq.(\ref{eigen3-2}) shows that a quantum doublet- this time robust
also against decoherence induced by its coupling with the plasmon
fields in the TLL leads- emerges in a tetrahedral JJN operated
near the FFP. Indeed, accounting for the fluctuations of the
plasmon fields amounts only to substitute $Y$ with the running
coupling strength, that is $\sim \left( \frac{L}{L_0}
\right)^{-\frac{g}{3}}$. Thus, we find that, via renormalization
of the boundary coupling strength, the collective plasmon modes
renormalize the instanton fugacity so as to make the gap $\Delta$
between the quantum doublet, $ | - Y \rangle_{1,2}$, and the first
excited singlet, $ | 2Y \rangle$, scale like $ \Delta ( L )  =
\Delta_0 \left( \frac{L}{L_0} \right)^{ - \frac{g}{3}}$. Thus, the
interaction with the plasmon modes enforces the gap between the
quantum doublet and the first excited state.

\subsection{Manipulation of the quantum doublet at the FFP}
\label{emerge2}

Let us, firstly, assume that the external phases $\vec{\mu}$ are
tuned nearby, but not exactly at, the triple degeneracy point,
$\mu_1^* , \mu_2^*$, that is,  $ | \mu_1 - \mu_1^* | / \pi \ll 1 ,
| \mu_2 - \mu_2^* | / \pi \ll 1$. The low-energy effective
Hamiltonian $\h_{\mathcal{F}}$ in the basis $ | R \rangle , | G
\rangle , | B \rangle$ is now modified to

\begin{equation}
\h_{\mathcal{F}} [ \vec{\epsilon} ] = \left\{ \frac{\pi g u}{L} \left[ \frac{1}{9}
+ \frac{\vec{\epsilon}^2}{4 \pi^2}\right]
\right\} {\bf I} + \begin{bmatrix} - \frac{2}{3} \frac{\pi g u}{L} \frac{\epsilon_1}{2 \pi} &
                           - Y e^{ i \frac{\pi}{3}}  & - Y e^{ - i \frac{\pi}{3}}  \\ - Y e^{  -  i \frac{\pi}{3}}
 & \frac{\pi g u}{L} \left[ \frac{1}{3} \frac{\epsilon_1}{2 \pi} -
 \frac{\epsilon_2}{\sqrt{3} \pi} \right] & - Y e^{  i \frac{\pi}{3}}  \\ - Y e^{  i \frac{\pi}{3}} &
- Y e^{ - i \frac{\pi}{3}} & \frac{\pi g u}{L} \left[ \frac{1}{3} \frac{\epsilon_1}{2 \pi} +
 \frac{\epsilon_2}{\sqrt{3} \pi} \right]
                          \end{bmatrix}
\:\:\:\: ,
\label{cmxx1}
\end{equation}
\noindent where we have set $\epsilon_i = \mu_i - \mu_i^* ,
i=1,2$. Of course, for $\epsilon_1 = \epsilon_2 = 0$,
$\h_{\mathcal{F}}[ \vec{\epsilon} ]$ reduces back to the
Hamiltonian in Eq.(\ref{cmx1}). In order to rewrite $\h_{\mathcal{F}} [
\vec{\epsilon} ] $ using the states reported in Eq.(\ref{eigen1}),
one has to rotate it to $\bar{H}_{\mathcal{F}}[ \vec{\epsilon} ] =
P^\dagger \h_{\mathcal{F}} [ \vec{\epsilon} ]P$ obtaining

\begin{equation}
P^\dagger \h_{\mathcal{F}} [ \vec{\epsilon} ]P = \left\{ \frac{\pi g u}{L} \left[ \frac{1}{9}
+ \frac{\vec{\epsilon}^2}{4 \pi^2} \right]
\right\} {\bf I} + \begin{bmatrix} - Y &  a - i\frac{b}{\sqrt{3}} & a + i\frac{b}{\sqrt{3}} \\
 a + i\frac{b}{\sqrt{3}} & - Y  &  a - i\frac{b}{\sqrt{3}} \\ a - i\frac{b}{\sqrt{3}}  &  a + i\frac{b}{\sqrt{3}} & 2 Y
                                                                      \end{bmatrix}
\:\:\:\: ,
\label{eigen3-1}
\end{equation}
\noindent with $ a = -\frac{1}{3}\frac{\pi g u}{L}
\frac{\epsilon_1}{2 \pi} , b = -\frac{\pi g u}{L}
\frac{\epsilon_2}{\sqrt{3}\pi} $. From Eq.(\ref{eigen3-1}), by
simply keeping the matrix elements of $P^\dagger \h_{\mathcal{F}}
[ \vec{\epsilon} ]P$ involving the low energy twofold degenerate
ground state, one has

\begin{equation}
\h_{\rm Eff} = \left\{ \frac{\pi g u}{L} \left[ \frac{1}{9}
+ \frac{\vec{\epsilon}^2}{4 \pi^2} \right] - Y
\right\} {\bf I}_2+ \begin{bmatrix}  0 &  a -i \frac{b}{\sqrt{3}} \\
 a + i \frac{b}{\sqrt{3}} &  0
                          \end{bmatrix}
\:\:\:\: .
\label{eigen4}
\end{equation}
\noindent From Eq.(\ref{eigen4}), one sees that, apart from the
term proportional to the identity matrix,  the matrix $\h_{\rm
Eff}$ can be rewritten as $\h_{\rm Eff} = b_x \sigma^x + b_y
\sigma^y$, with $b_x = a$, and $b_y =  \frac{  b}{\sqrt{3}}$.

To add the $z$-component to the effective magnetic field one needs
to break the degeneracy between the $ | \Uparrow \rangle$- and the
$ | \Downarrow \rangle$-states of ${\bf S}_G$. This may be
realized by turning on a nonzero $B_\parallel$ - see
Eq.(\ref{parallel})- and amounts to introduce an additional
contribution to $\h_{\mathcal{F}}$ given (see \ref{appe3}) by

\begin{equation}
\delta \h_{\mathcal{F}} = - \frac{B_\parallel}{4} Y \begin{bmatrix} 0 & z & z^* \\
z^*     &  0    & z  \\ z &
z^* & 0
                       \end{bmatrix}
\:\:\:\: ,
\label{eigen5}
\end{equation}
\noindent with $z = \frac{\pi}{3} e^{-i \frac{\pi}{6}}$. When
transforming to the states given in Eq.(\ref{eigen1}), one obtains

\begin{equation}
 P^\dagger \delta \h_{\mathcal{F}} P = - \frac{\sqrt{3}}{4} B_\parallel Y \begin{bmatrix}
                                                        1 & 0 & 0\\
                                                        0 & -1 & 0\\
                                                        0 & 0 & 0
                                                       \end{bmatrix}
\:\:\:\: .
\label{eigen6}
\end{equation}
\noindent From Eq.(\ref{eigen6}) one readily gets the low energy
effective two states Hamiltonian

\begin{equation}
\h_{\rm Eff} [ \vec{b} ]  = \left\{ \frac{\pi g u}{L} \left[ \frac{1}{9}
+ \frac{\vec{\epsilon}^2}{4 \pi^2}\right] - Y
\right\} {\bf I}_2+ \vec{b} \cdot \vec{\sigma}
\;\;\;\; ,
\label{eigen7}
\end{equation}
\noindent with $b_x = a$, $ b_y = \frac{b}{\sqrt{3}} $, and $b_z =
- \frac{\sqrt{3}}{4} B_\parallel Y $. The Hamiltonian $\h_{\rm
Eff} [ \vec{b} ] $ is then the Hamiltonian for a spin in an
external magnetic field, whose components may be manipulated by
acting on the external control parameters of the tetrahedral JJN.

For instance, applying a modulation in time to the phases
$\vec{\mu}$ and to the flux $f$, one may change the relative sign
between the two states according to the procedure outlined in
Ref.\cite{blatter1}. Indeed, one may modulate in time
$\epsilon_2$, so that $\epsilon_2 ( t ) = \nu \sin ( \omega_0 t
)$. This results in an effective $\vec{b}$ field given by $\vec{b}
= ( b_x , \tilde{b} \sin ( \omega_0 t ) , b_z)$, with $b_x , b_z $
constant and $ \tilde{b} = \frac{\pi g u}{3 L} \nu $. The
instantaneous eigenvalues of
 $ \h_{\rm Eff} [ \vec{b} ( t ) ]$ are then given by $\pm \Lambda (t ) = \pm \sqrt{b_x^2 + b_z^2 + \tilde{b}^2 \sin^2
( \omega_0 t )}$, while the corresponding adiabatic eigenstates
are given by

\begin{align}
| - \Lambda ( t ) \rangle &= e^{ i \xi_- ( t ) } \left\{
\cos \left( \frac{\Theta ( t ) }{2} \right)  | - Y \rangle_1
+ \sin \left( \frac{\Theta ( t ) }{2} \right) e^{ - i \Phi ( t ) } | - Y \rangle_2
\right\} \nonumber \\
| \Lambda ( t ) \rangle &= e^{ i \xi_+ ( t ) } \left\{ -
\sin \left( \frac{\Theta ( t ) }{2} \right)  | - Y \rangle_1
+ \cos \left( \frac{\Theta ( t ) }{2} \right) e^{ - i \Phi ( t ) } | - Y \rangle_2
\right\}
\:\:\:\: ,
\label{adiabei1}
\end{align}
\noindent with $ \cos ( \Theta ) = - b_z / \Lambda ( t ) $ and
$\Phi ( t ) = \text{arg} ( - b_x - i \tilde{b} \sin ( \omega_0 t
))$. As usual, the phases $\xi_\pm ( t )$ are chosen so as to
satisfy the condition
\begin{equation}
\langle \Lambda ( t ) | \frac{d}{d t } | \Lambda ( t ) \rangle =
\langle - \Lambda ( t ) | \frac{d}{d t } | - \Lambda ( t ) \rangle = 0
\:\:\:\: .
\label{adiabei2}
\end{equation}
\noindent Preparing, at $t=0$,  the system in the state $- |
\Lambda ( 0 )  \rangle$, after twice a period $2 T = 4 \pi
\omega_0$, the relative phases between $| - Y \rangle_1$ and $| -
Y \rangle_2$ becomes $\Delta \Phi = - \frac{1}{2} \int_0^{ 4\pi /
\omega_0} \: d t \: \dot{\Phi} ( t ) [ 1 - \cos ( \Theta ( t ) )
]$. Setting $b_z = 0$ and $\nu = b_x$, one finds $\Delta \Phi =
\pi$ and, thus, the relative sign between the two states is
exchanged. This procedure allows for engineering a NOT port,
using the quantum doublet emerging in a tetrahedral JJN at the
FFP.

To read the state of the quantum doublet one may look at the
pattern of the Josephson currents arising circulating in the JJN
when it is biased off the triple-degeneracy point. Indeed, within
the restricted subspace ${\mathcal F}$, the current operators are
given by

\begin{align}
I_{1,{\mathcal F}} &= e^* \frac{\partial P^\dagger \h_{\mathcal{F}} [ \vec{\epsilon} ]P}{\partial \epsilon_1} =
\frac{e^* g u}{2 \pi L } \epsilon_1 {\bf I} - \frac{e^* gu}{6 L}
\begin{bmatrix} 0 & 1 & 1 \\ 1 & 0 & 1 \\
 1 & 1 & 0
\end{bmatrix}  \nonumber \\
I_{2,{\mathcal F} }&= e^* \frac{\partial P^\dagger \h_{\mathcal{F}} [ \vec{\epsilon} ]P}{\partial \epsilon_2} =
\frac{e^* g u}{2 \pi L } \epsilon_2 {\bf I} + i \frac{e^* gu}{3 L}
\begin{bmatrix} 0 & 1 & -1 \\ -1 & 0 & 1 \\
 1 & -1 & 0
\end{bmatrix}
\:\:\:\: .
\label{currest}
\end{align}
\noindent From Eq.(\ref{currest}), one obtains
\begin{align}
  ~_1\langle - Y | I_{1,{\mathcal F}} | - Y \rangle_1 &= \frac{e^* gu}{2 \pi L} \epsilon_1 =\!\! ~_2\langle - Y | I_{1,{\mathcal
F}} | - Y \rangle_2 \\
  ~_1\langle - Y | I_{1,{\mathcal F}} | - Y \rangle_2 &= -\frac{e^* gu}{6 L} \\
  ~_1\langle - Y | I_{2,{\mathcal F}} | - Y \rangle_1 &= \frac{e^* gu}{2 \pi L} \epsilon_2 =\!\! ~_2\langle - Y | I_{2,{\mathcal
F}} | - Y \rangle_2 \\
  ~_1\langle - Y | I_{2,{\mathcal F}} | - Y \rangle_2 &= i\frac{e^* gu}{3 L}
\label{avecur1}
\end{align}
\noindent On the generic state of the doublet given by
$\left\lvert \alpha\right\rangle=\cos\frac{\theta}{2}\left\lvert
-Y\right\rangle_1+e^{i\phi}\sin\frac{\theta}{2}\left\lvert
-Y\right\rangle_2$, the expectation value of the current operators
are
\begin{align}
  \left\langle \alpha\right\rvert I_{1,{\mathcal F}} \left\lvert \alpha\right\rangle &=-\frac{e^* gu}{6L} \frac{1}{2}\langle \sigma_x \rangle_\alpha+ \frac{e^* gu}{2\pi L}\epsilon_1 \notag\\
  \left\langle \alpha\right\rvert I_{2,{\mathcal F}} \left\lvert \alpha\right\rangle &=-\frac{e^* gu}{3L} \frac{1}{2}\langle \sigma_y \rangle_\alpha+ \frac{e^* gu}{2\pi L}\epsilon_2
  \label{eq:avecur2}
\end{align}
\noindent From Eqs.(\ref{stab12.b}) and Eq.(\ref{eq:avecur2}), it
is easy to determine the current pattern identifying each
degenerate state $| - Y \rangle_1 , | - Y \rangle_2 $.

\section{Engineering a tetrahedral JJN operating near the FFP}
\label{engineer0}

Spinless TLL leads may be realized also with classical Josephson
junctions - i.e. using junctions for which $E_J / E_c \geq 1$-
\cite{chakra}, which may be easily and reliably fabricated with
well tested technologies \cite{haviland}. In this realization the
Luttinger parameter $g$ is given by $g \sim \sqrt{ \pi E_J / E_c}$
\cite{chakra} and a tetrahedral JJN operating near the FFP may be
fabricated by requiring that $g=2$ and then setting $L_* \sim
10^3$; this since, for $L \geq L_*$ and $1<g<3$, the phase slip
operators destabilize the SFP. The requirement $g=2$ may be easily
satisfied by using junctions for which $E_J / E_c \sim 1.3$.
Choosing $\lambda \sim  E_J / 3$ (see Eq.\eqref{eq:couplings}) one
gets that $ L_* \sim \left( \frac{E_J}{E_3}
\right)^{\frac{g}{g-1}} \: L_0 $ is of order $ L_* \approx 10^2
L_0$ where $L_0$ is the reference length of a chain with
parameters $E_J , E_3$. Setting $L_0 \sim 10$ yields then $L_*
\sim 10^3$.

Flux noise induced by the shift $\{ f \to f + \delta_i \}$ affects
the stability of the quantum doublet since its effects amount to
introduce an effective interaction between ${\bf S}_G$ and a
magnetic field ${\bf B}$ breaking the degeneracy between the $ | R
\rangle , | G \rangle$ and $ | B \rangle$-states. As shown in
subsection \ref{strong0}, these fluctuations are potentially
dangerous, as they may induce an effective ${\bf B}$-field acting
on the spin ${\bf S}_G$, which breaks the degeneracy between $ | R
\rangle , | G \rangle$ and $ | B \rangle$-states. Since the
running coupling strength associated to $B_\perp$ scales with $L$
as $b_\perp ( L ) = L B_\perp$ one sees that the scale- $L_{\rm
Stop}$- at which " dangerous" instanton trajectories will be
suppressed by $B_\perp$ may be be defined by the condition $
b_\perp ( L_{\rm Stop} ) \sim y ( L_{\rm Stop} )$, from which one
gets $L_{\rm Stop} \sim \left( \frac{Y}{B_\perp}
\right)^{\frac{3}{g}} L_0$.

To provide a rough estimate of $L_{\rm Stop}$, one may approximate
the actual instanton as a double-well instanton, by fitting the
parameters of the double-well potential $V_{\rm DB}$ so that the
minimum and the maximum points (and the values of $V_{\rm DB}$ at
the corresponding points) coincide with the ones obtained from
${\mathcal H}_{\bf B}$. A standard computation
\cite{scoleman,glhek} allows then to estimate the instanton
fugacity as

\begin{equation}
Y \approx 3 \pi \sqrt{\frac{E_3 \pi u}{g L}}\exp \left[ - 4.36 \sqrt{\frac{g L E_3}{3 \pi u}}\right]
\:\:\:\: .
\label{instamp}
\end{equation}
\noindent Using the same fabrication parameters as before and
fixing $ | \delta_i | \sim \pi /20$ (i.e., $ B_\perp \sim 0.2 E_J
( \delta_i )^2$), one gets that $ L_{\rm Stop} \sim 8\times 10^3$.
As a result, one may infer that noise in the external flux $f$
(described in our approach by $B_\perp \neq 0$ and by a finite
$L_{\rm Stop} $) does not affect the quantum doublet provided that
$L_{\rm Stop}
> L_*$.

\section{Concluding Remarks}
\label{conclusions}

In this paper we analyzed the phases accessible to a tetrahedral
JJN made by coupling a tetrahedral qubit \cite{blatter1,blatter2}
to three JJ chains acting as TLL leads.

We showed that, in a pertinent range of the fabrication and
control parameters, a robust attractive FFP emerges due to the
geometry of the tetrahedral JJN . In our approach the central
region - made by the tetrahedral qubit- is treated as a quantum
impurity of this low dimensional network. As a result, the
emergence of a FFP is a non perturbative phenomenon arising from
the strong coupling of the impurity with the TLL superconducting
leads.

We argued that the new stable FFP is associated with the emergence
of a doubly degenerate ground state, which may be regarded as a
quantum doublet described by a spin $1/2$ degree of freedom,
coupled to the plasmon modes of the superconducting TLL leads via
the boundary interaction. We showed that this quantum doublet is
robust not only against the noise in the external control
parameters (magnetic flux, gate voltage) but also against the
decoherence induced by the coupling of the tetrahedral qubit with
the superconducting leads. For this purpose, we showed that, as
the network size increases, the instanton operators, arising from
the interaction of the central region with the plasmon modes of
the leads, contribute to enforce the energy gap between the
twofold degenerate ground state and the first excited state.

We also pointed out how one may device protocols allowing to read
and manipulate the state of the quantum doublet emerging at the
FFP; we saw that this may be achieved by connecting the
tetrahedral JJN to three bulk superconductors at fixed phases $\{
\varphi_j \}$ ($j=1,2,3$ - see Fig.\ref{fig:device}). Indeed, we
showed that, acting on the $\{ \varphi_j \}$, induces an
``effective magnetic field'', which couples to the emerging
two-level quantum system, providing a tool to prepare the
two-level quantum system in a given state.

Finally, it is worth to point out that superconducting devices
such as the tetrahedral Josephson junction network analyzed in
this paper may be used to simulate physical behaviors realizable
in Kondo systems. Indeed, our RG analysis showed that, for $g=1$
and $G_z ( l = 0  ) =0$, the tetrahedral JJN   may be used to
simulate a Kondo spin pertinently coupled to band electrons.

\section{Acknowledgements}

We thank Ian Affleck and Alioscia Hamma for enlightening
discussions at various stages of our research. We benefited from
conversations with Maria Cristina Diamantini, Gianluca Grignani,
Mario Rasetti and Andrea Trombettoni. This work has been partly
supported by INFN.

\appendix

\section{The central region energy eigenstates}
\label{appe1}

In this appendix we report the full spectrum of the Hamiltonian
given in Eq.(\ref{tetra1}), for a generic value of the applied
flux $f$. In particular, we will single out the twofold degenerate
ground state whose levels have been used in section
\ref{device_2}, to define the effective spin-$1/2$ operator ${\bf
S}_G$.  The eigenstates, together with the corresponding energy
eigenvalues,  are given by \footnote{notice that the spin labels
correspond to sites 0,1,2,3, respectively; we shall set
$s=\sqrt{1+3\sin^2 f}$ and $t=\sqrt{3+\cos^2 f}$ henceforth;
moreover  we shall label the energy eigenstates by means of two
quantum numbers: the former ones refer to the total spin momentum
of the states, the latter ones to the $z$-component of the total
spin}.:

 $\boldsymbol{m=2}$: a fully polarized spin-2 state:
    \begin{equation}
      \left\lvert 2\right\rangle = \left\lvert \upa\upa\upa\upa\right\rangle \,,\quad (\varepsilon_2=-2H)
      \label{eq:state2}
    \end{equation}
 $\boldsymbol{m=1}$: four states given by
    \begin{align}
      \left\lvert 1,1\right\rangle &= \frac{1}{\sqrt{2t(t-\cos f)}}\left[ (-\cos f +t) \left\lvert \dwa\upa\upa\upa\right\rangle
+ \left\lvert \upa\dwa\upa\upa\right\rangle +
      \left\lvert \upa\upa\dwa\upa\right\rangle + \left\lvert \upa\upa\upa\dwa\right\rangle\right] \nonumber \\
      \varepsilon_{1,1}(f) &=-H +\frac{E_J}{2}\left( -\cos f -\sqrt{3+\cos^2 f} \right) \label{eq:state1.1} \\
      \left\lvert 1,2\right\rangle &= \frac{1}{\sqrt{2t(t+\cos f)}}\left[ (-\cos f -t) \left\lvert \dwa\upa\upa\upa\right\rangle
+ \left\lvert \upa\dwa\upa\upa\right\rangle +
      \left\lvert \upa\upa\dwa\upa\right\rangle + \left\lvert \upa\upa\upa\dwa\right\rangle\right] \nonumber \\
      \varepsilon_{1,2}(f) &=-H +\frac{E_J}{2}\left( -\cos f +\sqrt{3+\cos^2 f} \right) \label{eq:state1.2} \\
      \left\lvert 1,3\right\rangle &=\frac{1}{\sqrt{3}}\left[ \left\lvert \upa\dwa\upa\upa\right\rangle
+e^{i\frac{2\pi}{3}}\left\lvert \upa\upa\dwa\upa\right\rangle +
      e^{-i\frac{2\pi}{3}} \left\lvert \upa\upa\upa\dwa\right\rangle\right] \nonumber \\
      \varepsilon_{1,3}(f)&=-H +\frac{E_J}{2}\left( \cos f -\sqrt{3}\sin f \right) \label{eq:state1.3} \\
      \left\lvert 1,4\right\rangle &=\frac{1}{\sqrt{3}}\left[ \left\lvert \upa\dwa\upa\upa\right\rangle +
e^{-i\frac{2\pi}{3}}\left\lvert \upa\upa\dwa\upa\right\rangle +
      e^{i\frac{2\pi}{3}} \left\lvert \upa\upa\upa\dwa\right\rangle\right] \nonumber \\
      \varepsilon_{1,4}(f)&=-H +\frac{E_J}{2}\left( \cos f +\sqrt{3}\sin f \right) \label{eq:state1.4}
    \end{align}
\noindent
   $\boldsymbol{m=0}$: six states given by
    \begin{align}
      \left\lvert 0,1\right\rangle &= \frac{1}{\sqrt{6s(s-\sqrt{3}\sin f)}}\biggl[ \left( \left\lvert
\dwa\dwa\upa\upa\right\rangle +
            e^{i\frac{2\pi}{3}}\left\lvert \dwa\upa\dwa\upa\right\rangle + e^{-i\frac{2\pi}{3}}\left\lvert
\dwa\upa\upa\dwa\right\rangle\right) \nonumber \\
 &+ \left( \sqrt{3}\sin f -s \right)\left( \left\lvert \upa\upa\dwa\dwa\right\rangle +
      e^{i\frac{2\pi}{3}}\left\lvert \upa\dwa\upa\dwa\right\rangle + e^{-i\frac{2\pi}{3}}\left\lvert
\upa\dwa\dwa\upa\right\rangle \right)\biggr] \nonumber \\
      \varepsilon_{0,1}(f)&=\frac{E_J}{2}\left( \cos f -\sqrt{1+3\sin^2 f} \right) \label{eq:state0.1} \\
      \left\lvert 0,2\right\rangle &= \frac{1}{\sqrt{6s(s-\sqrt{3}\sin f)}}\biggl[ \left( \sqrt{3}\sin f -s \right) \left(
\left\lvert \dwa\dwa\upa\upa\right\rangle +
          e^{-i\frac{2\pi}{3}}\left\lvert \dwa\upa\dwa\upa\right\rangle + e^{i\frac{2\pi}{3}}\left\lvert
\dwa\upa\upa\dwa\right\rangle\right) \nonumber \\
  &+ \left( \left\lvert \upa\upa\dwa\dwa\right\rangle +
      e^{-i\frac{2\pi}{3}}\left\lvert \upa\dwa\upa\dwa\right\rangle + e^{i\frac{2\pi}{3}}\left\lvert
\upa\dwa\dwa\upa\right\rangle \right)\biggr] \nonumber \\
      \varepsilon_{0,2}(f)&=\varepsilon_{0,1}(f) \label{eq:state0.2} \\
      \left\lvert 0,3\right\rangle &= \frac{1}{\sqrt{6s(s+\sqrt{3}\sin f)}}\biggl[ \left( \left\lvert
\dwa\dwa\upa\upa\right\rangle +
            e^{i\frac{2\pi}{3}}\left\lvert \dwa\upa\dwa\upa\right\rangle + e^{-i\frac{2\pi}{3}}\left\lvert
\dwa\upa\upa\dwa\right\rangle\right) \nonumber \\
 &+ \left( \sqrt{3}\sin f +s \right)\left( \left\lvert \upa\upa\dwa\dwa\right\rangle +
      e^{i\frac{2\pi}{3}}\left\lvert \upa\dwa\upa\dwa\right\rangle + e^{-i\frac{2\pi}{3}}\left\lvert
\upa\dwa\dwa\upa\right\rangle \right)\biggr] \nonumber \\
      \varepsilon_{0,3}(f)&= \frac{E_J}{2}\left( \cos f +\sqrt{1+3\sin^2 f} \right) \label{eq:state0.3} \\
      \left\lvert 0,4\right\rangle &= \frac{1}{\sqrt{6s(s+\sqrt{3}\sin f)}}\biggl[ \left( \sqrt{3}\sin f +s \right) \left(
\left\lvert \dwa\dwa\upa\upa\right\rangle +
          e^{-i\frac{2\pi}{3}}\left\lvert \dwa\upa\dwa\upa\right\rangle + e^{i\frac{2\pi}{3}}\left\lvert
\dwa\upa\upa\dwa\right\rangle\right) \nonumber \\
&+ \left( \left\lvert \upa\upa\dwa\dwa\right\rangle +
      e^{-i\frac{2\pi}{3}}\left\lvert \upa\dwa\upa\dwa\right\rangle + e^{i\frac{2\pi}{3}}\left\lvert
\upa\dwa\dwa\upa\right\rangle \right)\biggr] \nonumber \\
      \varepsilon_{0,4}(f) &= \varepsilon_{0,3}(f) \label{eq:state0.4} \\
      \left\lvert 0,5\right\rangle &= \frac{1}{\sqrt{6}}\biggl[ \left( \left\lvert \dwa\dwa\upa\upa\right\rangle + \left\lvert
\dwa\upa\dwa\upa\right\rangle +
            \left\lvert \dwa\upa\upa\dwa\right\rangle\right) - \left( \left\lvert \upa\upa\dwa\dwa\right\rangle + \left\lvert
\upa\dwa\upa\dwa\right\rangle +
      \left\lvert \upa\dwa\dwa\upa\right\rangle\right)\biggr] \nonumber \\
     \varepsilon_{0,5}(f)&=-E_J\left( \cos f -1 \right) \label{eq:state0.5} \\1
      \left\lvert 0,6\right\rangle &= \frac{1}{\sqrt{6}}\left[ \left( \left\lvert \dwa\dwa\upa\upa\right\rangle + \left\lvert
\dwa\upa\dwa\upa\right\rangle +
            \left\lvert \dwa\upa\upa\dwa\right\rangle\right) + \left( \left\lvert \upa\upa\dwa\dwa\right\rangle + \left\lvert
\upa\dwa\upa\dwa\right\rangle +
      \left\lvert \upa\dwa\dwa\upa\right\rangle\right)\right] \nonumber \\
      \varepsilon_{0,6}(f)&=-E_J\left( \cos f +1 \right) \label{eq:state0.6}
\:\:\:\:.
    \end{align}
\noindent
   $\boldsymbol{m=-1}$: four states given by
    \begin{align}
      \left\lvert -1,1\right\rangle &= \frac{1}{\sqrt{2t(t-\cos f)}}\left[ (-\cos f +t) \left\lvert \upa\dwa\dwa\dwa\right\rangle
+ \left\lvert \dwa\upa\dwa\dwa\right\rangle +
      \left\lvert \dwa\dwa\upa\dwa\right\rangle + \left\lvert \dwa\dwa\dwa\upa\right\rangle\right] \nonumber \\
      \varepsilon_{-1,1}(f)&=H +\frac{E_J}{2}\left( -\cos f -\sqrt{3+\cos^2 f} \right) \label{eq:state-1.1} \\
      \left\lvert -1,2\right\rangle &= \frac{1}{\sqrt{2t(t+\cos f)}}\left[ (-\cos f -t) \left\lvert
\upa\dwa\dwa\dwa\right\rangle + \left\lvert \dwa\upa\dwa\dwa\right\rangle +
      \left\lvert \dwa\dwa\upa\dwa\right\rangle + \left\lvert \dwa\dwa\dwa\upa\right\rangle\right] \nonumber \\
      \varepsilon_{-1,2}(f) &= H +\frac{E_J}{2}\left( -\cos f +\sqrt{3+\cos^2 f} \right) \label{eq:state-1.2} \\
      \left\lvert -1,3\right\rangle &=\frac{1}{\sqrt{3}}\left[ \left\lvert \dwa\upa\dwa\dwa\right\rangle +
e^{-i\frac{2\pi}{3}}\left\lvert \dwa\dwa\upa\dwa\right\rangle +
      e^{i\frac{2\pi}{3}} \left\lvert \dwa\dwa\dwa\upa\right\rangle\right] \nonumber \\
      \varepsilon_{-1,3}(f)&=H +\frac{E_J}{2}\left( \cos f -\sqrt{3}\sin f \right) \label{eq:state-1.3} \\
      \left\lvert -1,4\right\rangle &=\frac{1}{\sqrt{3}}\left[ \left\lvert \dwa\upa\dwa\dwa\right\rangle +
e^{i\frac{2\pi}{3}}\left\lvert \dwa\dwa\upa\dwa\right\rangle +
      e^{-i\frac{2\pi}{3}} \left\lvert \dwa\dwa\dwa\upa\right\rangle\right] \nonumber \\
      \varepsilon_{-1,4}(f) &= H +\frac{E_J}{2}\left( \cos f +\sqrt{3}\sin f \right) \label{eq:state-1.4}
\:\:\:\: .
    \end{align}
  $\boldsymbol{m=-2}$: this is again a fully polarized spin state
  given by
    \begin{equation}
      \left\lvert -2\right\rangle = \left\lvert \dwa\dwa\dwa\dwa\right\rangle \,,\quad (\varepsilon_{-2}=2H)
      \label{eq:state-2}
    \end{equation}
\noindent From the knowledge of these states the effective
Hamiltonian for the central region ${\bf T}$ given in section
\ref{device_1} may be easily derived.

\section{DC-Josephson current pattern at weak couplings}
\label{weak1}

To induce a DC-Josephson current pattern across the JJN, one
has to apply static phase differences to the end point of the network. Thus,
the currents may be easily computed   within imaginary time
path integral formalism discussed in section \ref{weak2}. Indeed,
if at the endpoint of branch $i$ a static phase $\varphi_i$ is applied,
the currents $I_1, I_2, I_3$ flowing across the three branches of the
JJN may be  computed by taking the logarithmic derivatives of
the partition function $\Z$ in Eq.(\ref{partf1})  with respect to
 $\mu_1 , \mu_2$  \cite{giusonew},
and are given by

\begin{align}
I_1 &= \frac{- e^*}{\beta} \left\{ \sqrt{2} \frac{\partial \ln{\mathcal Z}}{\partial \mu_1} -
\sqrt{6} \frac{\partial \ln{\mathcal Z}}{\partial \mu_2} \right\} \nonumber \\
I_2 &=  \frac{- e^*}{\beta} \left\{ - \sqrt{2} \frac{\partial \ln{\mathcal Z}}{\partial \mu_1} -
\sqrt{6} \frac{\partial \ln{\mathcal Z}}{\partial \mu_2} \right\}  \nonumber \\
I_3 &= \frac{- e^*}{\beta} \sqrt{6} \frac{\partial \ln{\mathcal Z}}{\partial \mu_2}
\;\;\;\; .
\label{stab12}
\end{align}
\noindent
To compute ${\mathcal Z}$, one has to sum over the oscillating
modes of the fields $\chi_j$, by pertinently taking into
account boundary conditions at both boundaries.
At the inner boundary ($x=0$),  these  are
determined by energy conservation and are given by

\begin{align}
\frac{ug}{2 \pi} \frac{\partial \vec{\chi} ( 0 , \tau ) }{\partial x}
=& 2 \bar{E}_1 \sum_j \vec{\alpha}_j \sin \left[ \vec{\alpha}_j
\cdot \vec{\chi} (0,\tau) \right]
  +4 \bar{E}_z \sum_j \vec{\alpha}_j {\bf S}_G^z \sin \left[ \vec{\alpha}_j
\cdot \vec{\chi} (0,\tau) + \frac{\pi}{2} \right] +\nn\\
& +4 \bar{E}_3 \sum_j\vec{\alpha}_j  {\bf S}_G^x \cos \left[ \frac{2 \pi}{3}
(j-2 ) \right]  \sin \left[ \vec{\alpha}_j
\cdot \vec{\chi} (0,\tau) \right] +\nn\\
& +4 \bar{E}_3 \sum_j \vec{\alpha}_j {\bf S}_G^y \sin \left[ \frac{2 \pi}{3}
(j-2 ) \right]  \sin \left[ \vec{\alpha}_j
\cdot \vec{\chi} (0,\tau) \right]
\;\;\;\; ,
\label{boucon}
\end{align}
\noindent where, in order to account for normal ordering of
boundary interaction operators, the boundary interaction strengths
have been redefined as $\bar{E}_\ell = \left( \frac{a}{L}
\right)^\frac{1}{g} \: E_\ell $, ($ \ell = 1,z,3$)
\cite{giuso1,giuso4}. From Eq.(\ref{boucon}), one easily sees
that, at the WFP,  energy conservation requires Neumann boundary
conditions, for the plasmon fields at $x=0$ (i.e. $\frac{\partial
\chi_1 ( 0 , \tau)}{\partial x} = \frac{\partial \chi_2 ( 0 ,
\tau)}{\partial x} = 0$). As evidenced in section \ref{strong1},
at the outer boundary ($x=L$) $\chi_1, \chi_2$ obey Dirichlet
boundary conditions: $\chi_1 ( L , \tau ) = \mu_1 , \chi_2 ( L ,
\tau ) = \mu_2$. As a result, the mode expansion of the fields
$\chi_j ( x , \tau )$ is
\begin{align}
\chi_j (x,t) &=
\mu_j + \sqrt{\frac{2}{g}}\sum_{n \in {\bf Z}}
\cos\left[\frac{\pi}{L}\left(n+\frac{1}{2}\right)x\right]
\frac{\alpha_j(n)}{n+\frac{1}{2}}e^{-\frac{\pi}{L}\left(n+\frac{1}{2}\right)u \tau}\nn\\
&\equiv \mu_j + \phi_j ( x , \tau )
\;\;\;\; ,
\label{stab1}
\end{align}
\noindent where the oscillator modes $\alpha_i ( n )$ satisfy the
algebra
\begin{equation}
[ \alpha_i ( n ) , \alpha_j ( m ) ] = \delta_{i , j } \: \delta_{n+m-1,0} \left( n + \frac{1}{2}
\right)
\:\:\:\: .
\label{stab2}
\end{equation}
\noindent For our purposes it is convenient to define ``spin-$1/2$
current operators'', that is, operators acting on the
two-dimensional Hilbert space of ${\bf S}_G$ and giving the
correct value of the current, when ${\bf S}_G$ is averaged over,
as well. It is straightforward to see that these operators are
given by

\begin{align}
I_1 &= -\frac{e^*}{\beta} \left\{ \sqrt{2} \frac{\partial \ln{\mathcal Z}_S }{\partial \mu_1} -
\sqrt{6} \frac{\partial \ln{\mathcal Z}_S }{\partial \mu_2} \right\} \nonumber \\
I_2 &= -\frac{e^*}{\beta} \left\{ - \sqrt{2} \frac{\partial \ln{\mathcal Z}_S}{\partial \mu_1} -
\sqrt{6} \frac{\partial \ln{\mathcal Z}_S }{\partial \mu_2}\right\}
\nonumber \\
I_3 &= - \frac{e^*}{\beta} \sqrt{6} \frac{\partial \ln{\mathcal Z}_S }{\partial \mu_2}
\;\;\;\; ,
\label{stab12.b}
\end{align}
\noindent
with

\begin{equation}
{\mathcal Z}_S = \int \: \prod_{ i = 1,2} {\mathcal D} \chi_j \: e^{ - S_{\rm Lead}
-  S_\textbf{B}^{(I)} - S_S}
\:\:\:\: .
\label{wow1}
\end{equation}
\noindent
Relying over the weak coupling assumption, one may
sum over the plasmon modes $\chi_j$ within a mean-field like approach.
In particular, because of the mode expansion in Eq.(\ref{stab1}),
one gets

\begin{equation}
\langle \cos [ \vec{\alpha}_j \cdot \vec{\chi} ( \tau ) ] \rangle = \cos
[ \vec{\alpha}_j \cdot \vec{\mu} ] \;\;\; , \;\;
\langle \sin[ \vec{\alpha}_j \cdot \vec{\chi} ( \tau ) ] \rangle = \sin
[ \vec{\alpha}_j \cdot \vec{\mu} ]
\:\:\:\: .
\label{wow2}
\end{equation}
\noindent
Thus, one readily sees that, resorting to the mean-field
approximation amounts to
trade $\vec{\chi} ( \tau )$ for the applied phase differences
$\vec{\mu}$. As a result, one obtains

\begin{align}
\frac{- e^*}{\beta} \frac{ \partial \ln{\mathcal Z}_S }{\partial \vec{\mu}} &=
2  e^* \bar{E}_1 \sum_j \vec{\alpha}_j \sin \left[ \vec{\alpha}_j
\cdot \vec{\mu}  \right] +
  4  e^* \bar{E}_z \sum_j \vec{\alpha}_j {\bf S}^z_G \cos \left[ \vec{\alpha}_j
\cdot \vec{\mu}  \right] +\nn\\
& +4  e^* \bar{E}_3 \sum_j\vec{\alpha}_j  {\bf S}^x_G \cos \left( \frac{2 \pi}{3}
(j-2 ) \right)  \sin \left[ \vec{\alpha}_j
\cdot \vec{\mu}  \right] +\nn\\
& +4  e^* \bar{E}_3 \sum_j \vec{\alpha}_j {\bf S}^y_G \sin \left( \frac{2 \pi}{3}
(j-2 ) \right)  \sin \left[ \vec{\alpha}_j
\cdot \vec{\mu}  \right]
\;\;\;\; .
\label{tech5}
\end{align}
\noindent There are two possible ways of interpreting
Eq.(\ref{tech5}): on one hand, one may regard the applied phases
(and the induced currents) as a probe of the two-level state
(which may be set by acting upon it with the external fields
$B_\parallel , B_\perp$). For instance, assuming that the system
lies within either one of the eigenstates of ${\bf S}^z_G$,  $ |
\Uparrow \rangle , | \Downarrow \rangle$, and computing the
average values of the spin operators as outlined in \ref{appe2},
from   Eq.(\ref{tech5}) one gets
\begin{equation}
 \langle \frac{ \partial \ln{\mathcal Z}_S }{\partial \vec{\mu}}
\rangle_S  = 2 \sum_j \vec{\alpha}_j \{ \bar{E}_1 \sin \left[ \vec{\alpha}_j
\cdot \vec{\mu}  \right] \pm \bar{E}_z \cos \left[ \vec{\alpha}_j
\cdot \vec{\mu}  \right] \}
\;\;\;\; ,
\label{tech6}
\end{equation}
\noindent
where the average is computed over the spin coordinates.
  The corresponding current pattern may be
derived from Eq.(\ref{stab12.b}) and from
Eq.(\ref{tech6}): clearly, it discriminates between the
two states $ | \Uparrow \rangle , | \Downarrow \rangle$.   The same procedure may
be applied to  probing a generic state of ${\bf S}_G$, obviously
as long as the modification in the state induced by the application
of the phases $\vec{\mu}$ is negligible. On the other hand,
when no other fields are applied to ${\bf S}_G$ (that is,
when $B_\parallel = B_\perp = 0$),
the phases (and, of course, the currents) themselves
may be regarded as defining an effective applied field ${\bf B}$, whose
components are given by

\begin{align}
{\bf B}^x &= - 4 \bar{E}_3   \sum_j \cos \left[ \frac{2 \pi}{3} ( j - 2 ) \right]
\cos [ \vec{\alpha}_j \cdot \vec{\mu} ] \nonumber \\
{\bf B}^y &= - 4 \bar{E}_3   \sum_j \sin \left[ \frac{2 \pi}{3} ( j - 2 ) \right]
\cos [ \vec{\alpha}_j \cdot \vec{\mu} ] \nonumber \\
{\bf B}^z &= + 4 \bar{E}_z  \sum_j \sin [ \vec{\alpha}_j \cdot \vec{\mu} ]
\:\:\:\: .
\label{tech56.a}
\end{align}
\noindent
In this case the phases may be used to drive the state of the two-level
system, just as   a local magnetic field applied to
a true spin-$1/2$ variable.

\section{The imaginary time action for a quantum spin-$1/2$ variable}
\label{appe2}

In this appendix we shall review the derivation of the imaginary
time path integral formalism  for a quantum spin-$1/2$ variable
${\bf S}_G$ since it has been used to determine the instanton
phases at the SFP.

The starting point is the Euclidean action ${\mathcal Z}_{\rm
Spin}$ for a spin-$1/2$ degree of freedom in an external magnetic
field $\vec{\mathcal B} $, whose dynamics is described by the
Hamiltonian

\begin{equation}
\h_{\rm Spin} = - \vec{\mathcal B} \cdot {\bf S}_G
\:\:\:\: .
\label{appe2.1}
\end{equation}
\noindent A crucial step is the decomposition of the identity
${\bf I}$ in the basis of the coherent states $ | \Phi , \Theta
\rangle$ as

\begin{equation}
| \Phi , \Theta \rangle = e^{ i \Phi} \cos \left( \frac{\Theta}{2} \right) | \Uparrow \rangle
+ \sin \left( \frac{\Theta}{2} \right) | \Downarrow \rangle
\:\:\:\: ,
\label{stronx.1}
\end{equation}
\noindent The average values of the components of the spin-$1/2$
variable on the state  $ | \Phi , \Theta \rangle$, are then given
by

\[
\langle \Phi , \Theta | {\bf S}^x_G | \Phi , \Theta \rangle = \frac{1}{2} \cos ( \Phi )
\sin ( \Theta ) \;\;\; , \;\;
\langle \Phi , \Theta | {\bf S}^y_G | \Phi , \Theta \rangle = \frac{1}{2} \sin ( \Phi )
\sin ( \Theta )
\]
\noindent
\begin{equation}
\langle \Phi , \Theta | {\bf S}^z_G | \Phi , \Theta \rangle = \frac{1}{2} \cos ( \Theta )
\:\:\:\: .
\label{cohere1}
\end{equation}
\noindent
In particular, the decomposition of the identity  is given by
\begin{equation}
{\bf I} = \int \: d \Omega \: | \Phi , \Theta \rangle \langle \Phi , \Theta |
\:\:\:\: ,
\label{appe2.2}
\end{equation}
\noindent
with

\begin{equation}
 \int \: d \Omega \: \ldots = \frac{1}{4 \pi} \: \int_0^{2 \pi} d \Phi \int_0^{\pi} \:
\sin ( \Theta ) d \Theta \: \ldots
\:\:\:\: .
\label{appe2.3}
\end{equation}
\noindent Though coherent states form a complete set, they are not
orthogonal to each other; thus, in the imaginary time path
integral formulation \cite{auerbach}, one has to take into account
the overlap amplitude between two coherent states

\begin{equation}
\langle \Phi_1 , \Theta_1 | \Phi_2 , \Theta_2 \rangle = e^{ - i [ \Phi_1 - \Phi_2 ] }
\cos \left( \frac{\Theta_1}{2} \right) \cos \left( \frac{\Theta_2}{2} \right)
+ \sin \left( \frac{\Theta_1}{2} \right) \sin \left( \frac{\Theta_2}{2} \right)
\:\:\:\: .
\label{appe2.3.1}
\end{equation}
\noindent Taking into account Eq.(\ref{appe2.3.1}), one gets that
the amplitude for ${\bf S}_G$ to tunnel from the state $ | \Phi_0
, \Theta_0 \rangle$ to the state $ | \Phi_1 , \Theta_1 \rangle$ in
an (imaginary) time $\tau$ is given by

\[
\langle \Phi_1 , \Theta_1 | e^{ - \tau H [ {\bf S}_G ]} | \Phi_0 , \Theta_0 \rangle =
 \int \: {\mathcal D} \Omega ( \tau ) \: e^{ \left[ - \frac{i}{2}
\int_0^\beta \: d \tau \: \dot{\Phi} ( \tau )  ( 1 - \cos ( \Theta ( \tau ))
+ \int_0^\beta \: d \tau \:  \vec{\mathcal B} \cdot {\bf S}_G ( \tau ) \right] }
\]
\begin{equation}
\equiv  \int \: {\mathcal D} \Omega ( \tau ) \: \exp \left[ - S_E [ \Phi ,
\Theta \right] ]
\:\:\:\: ,
\label{appe2.4}
\end{equation}
\noindent where $ H [ {\bf S}]$ is the spin Hamiltonian, while we
have defined the polar angles as in Eqs.(\ref{cohere1}), and the
path integral has to be computed over imaginary time trajectories
satisfying the boundary conditions  $\Theta ( 0 ) = \Theta_0 ,
\Theta ( \tau  ) = \Theta_1$, and $\Phi ( 0 ) = \Phi_0 , \Phi (
\tau ) = \Phi_1$.

The one- and two-spin imaginary time correlation functions we used
in section \ref{weak0} may be then derived by means of a
saddle-point approximation. As an example, let us consider the
case in which the applied field $B$ is uniform and directed along
the $z$-axis, corresponding to Eq.(\ref{appe2.4}) for the
imaginary time amplitude. The saddle-point equations for the
functions $\Phi ( \tau ) , \Theta ( \tau )$ are given by:

\begin{align}
 0 &= \frac{\delta S_E [ \Phi ,
\Theta ] }{\delta \Theta ( \tau ) } = \frac{1}{2} \: \sin ( \Theta ( \tau ))
[ i \dot{\Phi} ( \tau ) + B ]  \nonumber \\
 0 &= \frac{\delta S_E [ \Phi ,
\Theta ] }{\delta \Phi ( \tau ) } = \frac{i}{2} \: \dot{\Theta} ( \tau ) \sin ( \Theta
( \tau ))
\:\:\:\: .
\label{appe2.5}
 \end{align}
\noindent
Eqs.(\ref{appe2.5}) imply

\begin{equation}
\Theta ( \tau ) = {\rm constant} = \Theta_0 = \Theta_1 \;\;\; , \;\;
\Phi ( \tau ) = i B \tau + \Phi_0
\:\:\:\: .
\label{appe2.6}
\end{equation}
\noindent From Eqs.(\ref{appe2.4},\ref{appe2.6}) one finds out
that, in the saddle-point approximation,

\begin{equation}
\langle \Phi_1 , \Theta_1 | e^{ - \tau H [ {\bf S}_G ]} | \Phi_0 , \Theta_0 \rangle =
\delta ( \Theta_0 - \Theta_1 ) \delta ( \Phi_f - \Phi_i - i B \tau )
\: e^{ \frac{B \tau}{2} }
\:\:\:\: .
\label{appe2.7}
\end{equation}
\noindent
From Eq.(\ref{appe2.7}) and from the identity

\begin{equation}
\langle \sigma | e^{ - \tau H } | \sigma^{'} \rangle =
\frac{1}{( 4 \pi )^2} \: \int_0^{2 \pi} \: d \Phi_0 \:
d \Phi_1 \: \int_0^\pi \: d \Theta_0 \: d \Theta_1 \: \sin ( \Theta_0 )
\sin ( \Theta_1 ) \:
\langle \Phi_1 , \Theta_1 | e^{ - \tau H [ {\bf S}_G ]} | \Phi_0 , \Theta_0 \rangle
\:\:\:\: ,
\label{appe2.8}
\end{equation}
\noindent with ${\bf S}^z_G | \sigma \rangle = \sigma | \sigma
\rangle$, one finds out (assuming $B > 0$) that

\begin{equation}
\langle \sigma | e^{ - \tau H } | \sigma^{'} \rangle =
e^{ - \frac{\sigma B }{2} }
\:\:\:\: .
\label{appe2.9}
\end{equation}
\noindent All the other imaginary time average values listed in
Eqs.(\ref{corref1},\ref{spinc2}) may be derived following a
similar approach. For instance, to compute the average value of a
component of ${\bf S}_G$, one may use

\begin{equation}
\langle {\bf S}_G^a ( \tau ) \rangle = \sum_\sigma \langle \sigma | e^{ \tau H} {\bf S}_G^a
e^{ - \tau H} | \sigma \rangle e^{ - \frac{ \sigma B}{2}} / \sum_\sigma  e^{ - \frac{ \sigma B}{2}}
\:\:\:\: ,
\label{appe2.10}
\end{equation}
\noindent and

\[
 \langle \sigma | e^{ \tau H} {\bf S}_G^a
e^{ - \tau H} | \sigma \rangle = \frac{1}{( 4 \pi )^2} \: \int_0^{2 \pi} \: d \Phi_a \:
d \Phi_b \: \int_0^\pi \: d \Theta_a \: d \Theta_b \: \sin ( \Theta_a)
\sin ( \Theta_b ) \: e^{ - \frac{\sigma B \tau}{2}}
\langle \sigma | \Phi_a , \Theta_a \rangle  \times
\]
\begin{equation}
\langle \Phi_a , \Theta_a | {\bf S}_G^a|
\Phi_b , \Theta_b  \rangle
 \langle \Phi_b , \Theta_b | \sigma \rangle  e^{  \frac{\sigma B \tau}{2}}
\:\:\:\: ,
\label{appe2.11}
\end{equation}
\noindent to get the results given in
Eqs.(\ref{corref1},\ref{spinc2}). When $B_\parallel \neq 0$ one
may use a similar analysis, provided one chooses the $z$ axis
directed along the direction of $\vec{B}$ and rotates the
components of ${\bf S}_G$ accordingly.

\section{Derivation of the modulus and phase for instanton trajectories}
\label{appe3}

In this appendix, we derive the modulus $Y$ and the phase $\gamma$
of the instanton tunnelling amplitudes.

First of all we recall that the instanton trajectory may be
regarded as the imaginary time evolution, $\vec{P} = \vec{P} (
\tau )$, of the zero mode contribution to
Eq.(ref{eq:mode-exp-strong}); it describes a tunnelling event
between nearest neighboring sites on the lattice of the minima.
The ``bulk'' Euclidean action for the field $\vec{\chi}$

\begin{equation}
S^{(0)} = \frac{g}{ 4 \pi} \: \int_0^\beta  \: d \tau \: \int_0^L \: d x \:
\left[ \frac{1}{u} \left( \frac{ \partial \vec{\chi}}{ \partial \tau}
\right)^2 + u \left( \frac{ \partial \vec{\chi}}{ \partial x}
\right)^2  \right]
\;\;\;\; ,
\label{appe2.5c}
\end{equation}
\noindent
yields

\begin{equation}
S^{(0)} = \frac{1}{2} \: \int_0^\beta \: d \tau \:
\left[ M \left( \frac{ d \vec{\chi} ( \tau )  }{ d \tau} \right)^2 +
M \omega^2 ( \vec{\chi} ( \tau ) - \vec{\xi} )^2 \right]
+  \ldots \equiv S^{(0)  }[ \vec{\chi} ( \tau ) ] + \ldots
\:\:\:\: ,
\label{appe2.6c}
\end{equation}
\noindent with $\vec{\chi} ( \tau ) = \vec{\chi} ( 0 , \tau )$, $M
= L g / ( 6 \pi u )$, and $ M \omega^2 = u g / ( 2 \pi L)$. The
ellipses in Eqs.(\ref{appe2.5c},\ref{appe2.6c}) corresponds to
interactions between instantons, mediated by oscillations of the
plasmon bulk modes, which do not affect the computation of the
phase $\gamma$ and will be neglected  henceforth. The coupling
between the $\vec{\chi}$-modes and the spin degrees of freedom
occurs via the boundary Hamiltonian in Eq.(\ref{eq:hb2}), which
may be presented as

\begin{equation}
\h_{\bf B}^{(I)} = {\mathcal B} [ \vec{\chi} ( \tau) ] \cdot
{\bf S}_G ( \tau ) + B_0 [ \vec{\chi} ( \tau)]
\;\;\;\; ,
\label{appe2.7c}
\end{equation}
\noindent
with
\begin{equation}
  \begin{aligned}
 B_0 [\vec{\chi} ( \tau) ]  &=  2 E_1 \: \sum_{j=1}^3 \: \cos [ \vec{\alpha}_j \cdot \vec{\chi} ( \tau ) ]
\;\;\;\; ,  \\
 {\mathcal B}_x  [\vec{\chi} ( \tau) ]  &= 2 E_3 \sum_{j=1}^3 \: \cos \left[ \frac{2 \pi}{3}
(j - 2 ) \right] \: \cos [ \vec{\alpha}_j \cdot \vec{\chi} ( \tau ) ]
\;\;\;\; , \\
  {\mathcal B}_y  [\vec{\chi} ( \tau) ]  &= 2 E_3 \sum_{j=1}^3 \: \sin \left[ \frac{2 \pi}{3}
(j - 2 ) \right] \: \cos [ \vec{\alpha}_j \cdot \vec{\chi} ( \tau ) ]
\;\;\;\; , \\
{\mathcal B}_z  [\vec{\chi} ( \tau) ]  &=  2 E_z \sum_{j=1}^3 \: \cos [ \vec{\alpha}_j \cdot \vec{\chi} ( \tau ) +\frac{\pi}{2}] + B_{\parallel}
\;\;\;\; .
\end{aligned}
\label{appe2.7d}
\end{equation}
\noindent Focusing on the ``threefold degenerate'' point obtained
when $ ( \mu_1^* . \mu_2^* ) = \left( \frac{4 \pi}{3} , 0
\right)$, one easily realizes that the instanton trajectories of
interest lie along the sides of the triangle whose vertices
coincide with the $R, G, B$ points defined by $n_{12} = n_{13} =
0$. At the $R, G, B$-vertices one gets

\begin{equation}
\vec{\chi}_R = 2 \pi ( 1 , 0 ) \;\;\; , \;\;
\vec{\chi}_G = 2 \pi \left( \frac{1}{2} , - \frac{1}{\sqrt{3}} \right) \;\;\; , \;\;
 \vec{\chi}_B = 2 \pi \left( \frac{1}{2} ,  \frac{1}{\sqrt{3}} \right)
\;\;\;\; .
\label{appe2.9c}
\end{equation}
\noindent Any instanton path runs between two of the points in the
$\vec{\chi}$-configuration space listed in Eq.(\ref{appe2.9c}).
Moreover, due to
\begin{equation} \vec{\alpha}_2 = {\mathcal R} \left( \frac{2
\pi}{3} \right) \vec{\alpha}_1 \;\;\; ,\;\; \vec{\alpha}_3 =
{\mathcal R} \left( \frac{4 \pi}{3} \right) \vec{\alpha}_1
\;\;\;\; , \label{esette}
\end{equation}
\noindent
with

\begin{equation}
{\mathcal R} ( \theta ) =  \begin{bmatrix} \cos ( \theta ) & \sin ( \theta ) \\
                              - \sin ( \theta ) & \cos ( \theta )
                             \end{bmatrix}
\:\:\:\: ,
\label{eotto}
\end{equation}
\noindent one sees that the path connecting $B$ to $R$ may be
obtained by acting with ${\mathcal R} \left( \frac{2 \pi}{3}
\right) $ on the one connecting $G$ to $B$; in addition, the path
connecting $R$ to $G$ may be obtained by acting with ${\mathcal R}
\left( \frac{4 \pi}{3} \right) $ on the path connecting $G$ to$B$.
As a result, it is enough to compute only one tunnelling
amplitude, for example the one between $G$ and $B$. To do so, let
us parameterize such an instanton path as $ ( \chi_1 ( \tau ) ,
\chi_2 ( \tau )) = \left( \pi , \frac{2\pi}{\sqrt{3}} \sigma (
\tau ) \right)$, with $\sigma ( 0 ) = - \frac{1}{2}$, and $\sigma
( \beta ) = \frac{1}{2}$. The corresponding Euclidean action is
then given by

\begin{equation}
S [ \sigma ] = \int_0^\beta \: d \tau \: \left\{ \frac{M \pi^2}{6}
[ ( \dot{\sigma} )^2 + \omega^2 \sigma^2 ] - 2 E_1 - 2 E_3
\sqrt{4 - 3 \sin^2 \left( \pi \sigma \right) } \right\}
\;\;\;\; .
\label{appe2.10c}
\end{equation}
\noindent
The term $\propto \sigma^2$ is basically constant, along the instanton path. Thus,
the actual value of $S [ \sigma ] $ may, in principle, be computed by determining
the zero-action solution in the ``inverted potential'', $\sigma_{\rm Inst} ( \tau )$, as

\begin{equation}
\frac{M \pi^2}{6}  \dot{\sigma}_{\rm Inst}^2 ( \tau )
\approx 2 E_3
\left( \sqrt{4 - 3 \sin^2 \left( \pi \sigma_{\rm Inst} ( \tau ) \right) } +1 \right)
\:\:\:\: ,
\label{appe2.11c}
\end{equation}
\noindent and, then, evaluating $\bar{S} = S [ \sigma_{\rm Inst}
]$. Since the fugacity $Y = e^{ - \bar{S} }$ is strongly
renormalized by the interaction with the collective plasmon modes
of the bulk, it eventually scales as $ Y ( L) \sim L^{-
\frac{g}{3} }$ and, thus, one gets that $e^{ - \bar{S} } \sim Y (
L = L_0 )$.

The tunnelling amplitudes have also a phase stemming from the
topological term due to the spin \cite{wilczek}. To derive it, one
may consider that the probability amplitude to remain in the same
state (for example R), is not only given by the
instanton/anti-instanton contributions,  but also by the loops
around the three degenerate states. The action of such a loop will
induce an extra topological term arising from the trajectory of
the spin state during the loop. For a generic spin state
$\left\lvert \alpha(\tau)\right\rangle =
\cos\frac{\theta(\tau)}{2}\left\lvert \Upa\right\rangle +
e^{i\phi(\tau)}\sin\frac{\theta(\tau)}{2} \left\lvert
\Dwa\right\rangle $ the topological term can be written as:
\begin{equation}
  S_{\rm Top} = \frac{i}{2}\int_{0}^{T} d\tau\: \dot{\phi}(\tau)\left( 1+\cos \theta(\tau) \right)\,.
  \label{eq:stop}
\end{equation}
Thus the Euclidean action for a loop will be given by:
\begin{equation}
  S_{\rm loop} [ \vec{\chi} ( \tau ) ; \phi ( \tau ) ; \theta ( \tau ) ] =
  S_{\rm Top} [ \phi , \theta ] + 3 S_{\rm inst}\,.
\label{sloop}
\end{equation}
Writing the spin states along the instanton paths and substituting
the instanton coordinates $\vec\chi(\tau)$ into Eq.(\ref{eq:hbm}),
one finds the corresponding lowest energy eigenstate. For small
$B_{\parallel}$, one obtains:
\begin{align}
  \left\lvert \alpha(\tau)\right\rangle = \cos\frac{\theta_f}{2}\left\lvert \Upa\right\rangle +
  e^{i\gamma(\tau)} \sin\frac{\theta_f}{2} \left\lvert \Dwa\right\rangle
  \label{eq:inst-spin}
\end{align}
where $\theta_f$ is defined in Eq.\eqref{eq:eigstathb}
and $\gamma(\tau)=\arg\left[ B_x[\vec\chi(\tau)] + I B_y[\vec\chi(\tau)] \right]$.
Since $\theta$ remains constant, the evaluation of $S_{\rm Top}$ yields
\begin{equation}
  S_{\rm Top}=i\pi(1+\cos\theta_f)\approx i\pi(1-\frac{B_{\parallel}}{4})\;\;\;\; ,
  \label{eq:stop2}
\end{equation}
that is, the phase contribution to the amplitude of loop
tunnelling. Assuming that the three instanton tunnelling are
equivalent, one naturally assign to each tunnelling amplitude  a
third of the total phase. Thus, the single instanton tunnelling
amplitude is given by:
\begin{equation}
  \mathcal{Y}=Y e^{i\frac{\pi}{3} - i\frac{B_{\parallel}}{12}}\;\;\;\; ,
\end{equation}
\noindent which is the result used in the paper.

\biboptions{sort}


\begin{thebibliography}{10}
\expandafter\ifx\csname url\endcsname\relax
  \def\url#1{\texttt{#1}}\fi
\expandafter\ifx\csname urlprefix\endcsname\relax\def\urlprefix{URL }\fi
\expandafter\ifx\csname href\endcsname\relax
  \def\href#1#2{#2} \def\path#1{#1}\fi

\bibitem{boundary}
I.~Affleck, Quantum impurity problems in condensed matter physics, in: Exact
  Methods in Low-dimensional Statistical Physics and Quantum Computing: Lecture
  Notes of the Les Houches Summer School: Volume 89, July 2008,
  arXiv:0809.3474, Oxford Univ Pr, 2010, p.~3.

\bibitem{beenakker1}
H.~van Houten, C.~Beenakker, Quantum point contacts, Physics Today 49~(7)
  (1996) 22--27.
\newblock \href {http://dx.doi.org/10.1063/1.881503}
  {\path{doi:10.1063/1.881503}}.

\bibitem{beenakker2}
C.~Beenakker, H.~van Houten, Quantum point contacts, in: W.~Kirk, M.~Reed
  (Eds.), Nanostructures and mesoscopic systems: proceedings of the
  international symposium, Santa Fe, New Mexico, May 20-24, 1991, Academic
  Press, 1992.

\bibitem{chamon}
C.~Chamon, M.~Oshikawa, I.~Affleck, Junctions of three quantum wires and the
  dissipative hofstadter model, Phys. Rev. Lett. 91 (2003) 206403.
\newblock \href {http://arxiv.org/abs/cond-mat/0305121}
  {\path{arXiv:cond-mat/0305121}}, \href
  {http://dx.doi.org/10.1103/PhysRevLett.91.206403}
  {\path{doi:10.1103/PhysRevLett.91.206403}}.

\bibitem{chamon2}
M.~Oshikawa, C.~Chamon, I.~Affleck, Junctions of three quantum wires, J. Stat.
  Mech. 0602 (2006) P008.
\newblock \href {http://arxiv.org/abs/cond-mat/0509675}
  {\path{arXiv:cond-mat/0509675}}.

\bibitem{giuso4}
D.~Giuliano, P.~Sodano, Y-junction of superconducting josephson chains, Nuclear
  Physics, Section B 811~(3) (2009) 395--419.
\newblock \href
  {http://dx.doi.org/http://dx.doi.org/10.1016/j.nuclphysb.2008.11.011}
  {\path{doi:http://dx.doi.org/10.1016/j.nuclphysb.2008.11.011}}.

\bibitem{cardy}
J.~Cardy, Boundary conformal field theory, Encyclopedia of mathematical
  physics, arXiv:hep-th/0411189v2.

\bibitem{giamarchi}
T.~Giamarchi, Quantum physics in one dimension, Oxford University Press, USA,
  2004.

\bibitem{tsvel}
A.~M. Tsvelick, P.~B. Wiegmann, Exact results in the theory of magnetic alloys,
  Advances in Physics 32 (1983) 453 -- 713.

\bibitem{schlott}
P.~Schlottmann, Some exact results for dilute mixed-valent and heavy-fermion
  systems, Physics Reports 181~(1-2) (1989) 1 -- 119.
\newblock \href {http://dx.doi.org/10.1016/0370-1573(89)90116-6}
  {\path{doi:10.1016/0370-1573(89)90116-6}}.

\bibitem{nozi}
P.~Nozi\`eres, A.~Blandin, Kondo effect in real metals, J. Phys. France 41~(3)
  (1980) 193--211.
\newblock \href {http://dx.doi.org/10.1051/jphys:01980004103019300}
  {\path{doi:10.1051/jphys:01980004103019300}}.

\bibitem{afflud}
I.~Affleck, A.~Ludwig, Exact conformal-field-theory results on the multichannel
  kondo effect: Single-fermion green’s function, self-energy, and
  resistivity, Physical Review B 48~(10) (1993) 7297--7321.

\bibitem{fendlud}
P.~Fendley, A.~Ludwig, H.~Saleur, Exact conductance through point contacts in
  the $\nu= 1/3$ fractional quantum hall effect, Physical review letters
  74~(15) (1995) 3005--3008.

\bibitem{chang}
A.~M. Chang, Chiral luttinger liquids at the fractional quantum hall edge, Rev.
  Mod. Phys. 75~(4) (2003) 1449--1505.
\newblock \href {http://dx.doi.org/10.1103/RevModPhys.75.1449}
  {\path{doi:10.1103/RevModPhys.75.1449}}.

\bibitem{kanefish1}
C.~L. Kane, M.~P.~A. Fisher, Transmission through barriers and resonant
  tunneling in an interacting one-dimensional electron gas, Physical Review B
  46~(23) (1992) 15233--15262.

\bibitem{kanefish2}
C.~L. Kane, M.~P.~A. Fisher, Transport in a one-channel luttinger liquid, Phys.
  Rev. Lett. 68~(8) (1992) 1220--1223.
\newblock \href {http://dx.doi.org/10.1103/PhysRevLett.68.1220}
  {\path{doi:10.1103/PhysRevLett.68.1220}}.

\bibitem{reyes}
S.~A. Reyes, A.~M. Tsvelik, Crossed spin-$1/2$ heisenberg chains as a quantum
  impurity problem, Phys. Rev. Lett. 95~(18) (2005) 186404.
\newblock \href {http://dx.doi.org/10.1103/PhysRevLett.95.186404}
  {\path{doi:10.1103/PhysRevLett.95.186404}}.

\bibitem{giuso1}
D.~Giuliano, P.~Sodano, Effective boundary field theory for a josephson
  junction chain with a weak link, Nucl. Phys. B 711~(3) (2005) 480--504.
\newblock \href
  {http://dx.doi.org/http://dx.doi.org/10.1016/j.nuclphysb.2005.01.037}
  {\path{doi:http://dx.doi.org/10.1016/j.nuclphysb.2005.01.037}}.

\bibitem{giuso2}
D.~Giuliano, P.~Sodano, Boundary field theory approach to the renormalization
  of squid devices, Nucl. Phys. B 770~(3) (2007) 332--370.
\newblock \href
  {http://dx.doi.org/http://dx.doi.org/10.1016/j.nuclphysb.2007.02.015}
  {\path{doi:http://dx.doi.org/10.1016/j.nuclphysb.2007.02.015}}.

\bibitem{giusoepl}
D.~Giuliano, P.~Sodano, Pairing of cooper pairs in a josephson junction network
  containing an impurity, Europhysics Letters 88 (2009) 17012.
\newblock \href
  {http://dx.doi.org/http://dx.doi.org/10.1209/0295-5075/88/17012}
  {\path{doi:http://dx.doi.org/10.1209/0295-5075/88/17012}}.

\bibitem{giusolast}
D.~Giuliano, P.~Sodano, Competing boundary interactions in a josephson junction
  network with an impurity, Nuclear Physics B 837~(3) (2010) 153 -- 185.
\newblock \href {http://dx.doi.org/10.1016/j.nuclphysb.2010.04.022}
  {\path{doi:10.1016/j.nuclphysb.2010.04.022}}.

\bibitem{glark}
L.~Glazman, A.~Larkin, New quantum phase in a one-dimensional josephson array,
  Phys. Rev. Lett. 79~(19) (1997) 3736--3739.
\newblock \href {http://dx.doi.org/doi:10.1103/PhysRevLett.79.3736}
  {\path{doi:doi:10.1103/PhysRevLett.79.3736}}.

\bibitem{glhek}
F.~Hekking, L.~Glazman, Quantum fluctuations in the equilibrium state of a thin
  superconducting loop, Phys. Rev. B 55~(10) (1997) 6551--6558.
\newblock \href {http://dx.doi.org/10.1103/PhysRevB.55.6551}
  {\path{doi:10.1103/PhysRevB.55.6551}}.

\bibitem{giusonew}
D.~Giuliano, P.~Sodano, Frustration of decoherence in y-shaped superconducting
  josephson networks, New Journal of Physics 10 (2008) 093023.
\newblock \href
  {http://dx.doi.org/http://dx.doi.org/10.1088/1367-2630/10/9/093023}
  {\path{doi:http://dx.doi.org/10.1088/1367-2630/10/9/093023}}.

\bibitem{shon0}
Y.~Makhlin, G.~Sch\"on, A.~Shnirman, Quantum-state engineering with
  josephson-junction devices, Rev. Mod. Phys. 73~(2) (2001) 357--400.
\newblock \href {http://dx.doi.org/10.1103/RevModPhys.73.357}
  {\path{doi:10.1103/RevModPhys.73.357}}.

\bibitem{blatter1}
M.~V. Feigel'man, L.~B. Ioffe, V.~B. Geshkenbein, P.~Dayal, G.~Blatter,
  Superconducting tetrahedral quantum bits, Phys. Rev. Lett. 92~(9) (2004)
  098301.
\newblock \href {http://dx.doi.org/10.1103/PhysRevLett.92.098301}
  {\path{doi:10.1103/PhysRevLett.92.098301}}.

\bibitem{blatter2}
M.~V. Feigel'man, L.~B. Ioffe, V.~B. Geshkenbein, P.~Dayal, G.~Blatter,
  Superconducting tetrahedral quantum bits: Emulation of a noise-resistant
  spin- $ 1 2 $ system, Phys. Rev. B 70~(22) (2004) 224524.
\newblock \href {http://dx.doi.org/10.1103/PhysRevB.70.224524}
  {\path{doi:10.1103/PhysRevB.70.224524}}.

\bibitem{Usma} R.~B. Usmanov, L.~B. Ioffe, Theoretical
investigation of a protected quantum bit in a small Josephson
junction array with tetrahedral symmetry, Phys. Rev. B 69~(21)
(2004) 214513. \newblock \href {http://dx.doi.org/10.1103/PhysRevB.69.214513}
  {\path{doi:10.1103/PhysRevB.69.214513}}.


\bibitem{wiring}
C.~van~der Wal, F.~Wilhelm, C.~Harmans, J.~Mooij, Engineering decoherence in
  josephson persistent-current qubits, The European Physical Journal B -
  Condensed Matter and Complex Systems 31 (2003) 111--124.
\newblock \href {http://dx.doi.org/10.1140/epjb/e2003-00015-9}
  {\path{doi:10.1140/epjb/e2003-00015-9}}.

\bibitem{shulz}
H.~J. Shulz, G.~Cuniberti, P.~Pieri, Fermi liquids and luttinger liquids, in:
  Morandi et~al.  \cite{MorandiSod2000field}.

\bibitem{yang}
C.~N. Yang, C.~P. Yang, One-dimensional chain of anisotropic spin-spin
  interactions. ii. properties of the ground-state energy per lattice site for
  an infinite system, Phys. Rev. 150~(1) (1966) 327--339.
\newblock \href {http://dx.doi.org/10.1103/PhysRev.150.327}
  {\path{doi:10.1103/PhysRev.150.327}}.

\bibitem{haviland}
P.~Agren, K.~Andersson, D.~B. Haviland, Kinetic inductance and coulomb blockade
  in one dimensional josephson junction arrays, Journal of Low Temperature
  Physics 124 (2001) 291--304.
\newblock \href {http://dx.doi.org/10.1023/A:1017594322332}
  {\path{doi:10.1023/A:1017594322332}}.

\bibitem{Cardy1996scaling}
J.~Cardy, Scaling and renormalization in statistical physics, Cambridge Univ
  Pr, 1996.

\bibitem{cardy0}
J.~Cardy, Conformal invariance and surface critical behavior, Nuclear Physics B
  240~(4) (1984) 514--532.

\bibitem{anisotropic}
G.~Yuval, P.~W. Anderson, Exact results for the kondo problem: One-body theory
  and extension to finite temperature, Phys. Rev. B 1~(4) (1970) 1522--1528.
\newblock \href {http://dx.doi.org/10.1103/PhysRevB.1.1522}
  {\path{doi:10.1103/PhysRevB.1.1522}}.

\bibitem{chakra}
P.~Goswami, S.~Chakravarty, Dissipation, topology, and quantum phase transition
  in a one-dimensional josephson junction array, Phys. Rev. B 73~(9) (2006)
  094516.
\newblock \href {http://dx.doi.org/10.1103/PhysRevB.73.094516}
  {\path{doi:10.1103/PhysRevB.73.094516}}.

\bibitem{scoleman}
S.~Coleman, Aspects of symmetry: selected Erice lectures of Sidney Coleman,
  Cambridge Univ Pr, 1988, Ch. Chap. 7, The Uses of Instantons.

\bibitem{auerbach}
A.~Auerbach, F.~Berruto, L.~Capriotti, Quantum magnetism approaches to strongly
  correlated electrons, in: Morandi et~al.  \cite{MorandiSod2000field}.

\bibitem{wilczek}
F.~Wilczek, A.~Shapere, Geometric phases in physics, World Scientific Pub Co
  Inc, 1989.

\bibitem{MorandiSod2000field}
G.~Morandi, P.~Sodano, V.~Tognetti, A.~Tagliacozzo (Eds.), Field Theories for
  Low-Dimensional Condensed Matter Systems: Spin Systems and Strongly
  Correlated Electrons, Springer, 2000.

\end{thebibliography}
\end{document}